\documentclass[a4paper,11pt]{article}

\usepackage{jheppub} 

\usepackage[T1]{fontenc} 

\title{\boldmath Aspects of Gravitational Wave/Particle Duality:  Bulk Torsion $\leftrightarrow$ Boundary Gravity Correspondence}

\author[]{Rohit K. Gupta,}
\author[]{Supriya Kar}
\author[]{and R. Nitish}

\affiliation[]{Department of Physics and Astrophysics, University of Delhi, India 110007}

\emailAdd{rkgupta1@physics.du.ac.in}
\emailAdd{skkar@physics.du.ac.in}
\emailAdd{nitish@physics.du.ac.in}

\abstract{
A geometric torsion (GT) underlying a $2$-form in a $(4$$+$$1)$-dimensional $U(1)$ gauge theory is revisited with a renewed perspective for a non-perturbation (NP) gravity in $d$$=$$4$. In the context we provide evidences to a holographic correspondence between a bulk  GT and a boundary NP gravity. Interestingly the Killing symmetries in General Relativity (GR) are shown to provide a subtle clue to the quantum gravity. The NP gravity is shown to incorporate a $(B_{2}$$ \wedge $$F_{2})$ coupling, sourced by a non-Newtonian potential, to an exact geometry in GR.  Remarkably the NP correction is identified as a mass dipole and is shown to be sourced by a propagating GT. A detailed analysis is performed in a bulk GT to show a modification to the precession of perihelion in a boundary NP gravity. The perspective of an electromagnetic (EM) wave in the bulk is investigated to reveal a spin $2$ (mass-less) quantum sourced by an apparent 2-form. A Goldstone scalar is absorbed by the apparent 2-form to  describe a massive $2$-form in the coulomb gauge. Alternately a Goldstone scalar together with a local degree of GT and 2-form is argued to govern a composite (mass-less) spin $2$ particle in  Lorentz gauge. Both the scenarios, further ensure a graviton in a boundary NP gravity.  A qualitative analysis reveals a (non-interacting) graviton underlying a plausible gravitational wave/particle duality in NP gravity. }

\begin{document} 
\maketitle
\flushbottom

\def\ra{{\rightarrow}}
\def\a{{\alpha}}
\def\b{{\beta}}
\def\l{{\lambda}}
\def\eps{{\epsilon}}
\def\T{{\Theta}}
\def\t{{\theta}}
\def\co{{\cal O}}
\def\car{{\cal R}}
\def\caf{{\cal F}}
\def\cs{{\Theta_S}}
\def\pr{{\partial}}
\def\tri{{\triangle}}
\def\na{{\nabla }}
\def\S{{\Sigma}}
\def\s{{\sigma}}
\def\sp{\vspace{.1in}}
\def\hs{\hspace{.25in}}

\newcommand{\be}{\begin{equation}} \newcommand{\ee}{\end{equation}}
\newcommand{\bea}{\begin{eqnarray}}\newcommand{\eea}
{\end{eqnarray}}

\section{Introduction}\label{section01}
Symmetries are powerful theoretical tools and their study quite often help to explore new physics. In particular the Killing symmetries
play a significant role in GR which is elegantly described by a metric $g_{\mu\nu}({\mathbf{x}},t)$ field in $(3$$+$$1)$-dimensions. Thus the isometries in GR ensure that the Lie derivative ${\cal L}_K g_{\mu\nu}$$=$$0$. They lead to Killing equations 
$\nabla_{(\mu}K_{\nu)}$$=$$0$ and their solution defines a Killing vector. Interestingly the 
conserved charges underlying each Killing vector is known to contribute to a gravitational potential \cite{weinberg-book}. However a potential in GR is not uniquely defined. For instance, all exact solutions in GR are defined with different gravitational potentials mostly underlying various isometries. These potentials are known to deform the geometries encoded in a line-element. 

\sp
\noindent
Generically a gravitational potential in GR describes the Newtonian gravity in an appropriate limit. In fact a static, $S_2$ symmetric, vacuum solution is precisely governed by the Newtonian potential underlying a 
scalar field $\phi$. The vacuum geometry ensures a non-trivial space-time curvature in conformal  tensor $C_{\mu\nu\lambda\rho}$. Similarly a static, $S_2$ symmetric, charged solution in GR coupled to an $U(1)$ gauge theory is governed by a rank one tensor field $A_{\mu}({\mathbf{x}},t)$ in addition to the scalar field. The gauge field consistently retains the Newtonian gravity via its equation of motion and ensures a non-vacuum solution in GR. 

\sp
\noindent
Along the line of thought a Ricci scalar ${\cal R}$ is known to couple to gauge theories underlying 
higher $p$-forms, $i.e.$ for $p$$=$$(2,3,4)$, in the bosonic sector of $d$$=$$10$ superstring effective action \cite{GSW}. However a classical description in the bosonic sector of a string effective action becomes sensible in an arbitrary dimension. For instance in $d$$=$$4$, a static, 
$S_2$ symmetric, charged string solution has been shown to describe a shrinking event horizon \cite{Garfinkle} which is otherwise not feasible in GR (coupled to an $U(1)$ theory). In principle GR can only couple to a $2$-form among all the higher $p$-forms which in turn may be viewed as a metric-(pseudo)scalar theory \cite{metric-scalar, Naruko-2016}. 

\sp
\noindent
In the context a constant of motion in GR, underlying the Killing vectors, is known to describe an one dimensional dynamical system on a equatorial plane. Interestingly an effective potential $V_{\rm eff}$, for  time-like geodesics with a vacuum geometry, is known to possess a non-Newtonian potential term in GR which in turn is believed to possess a clue to a quantum correction. It is known to vary as an inverse cube of the radial distance and may seen to be sourced by a formal combination $(\phi F_{\mu\nu})$ of a scalar field and an EM-field. A non-vanishing conserved force 
$\nabla^{\mu}(\phi F_{\mu\nu})$$=$$(\partial^{\mu}\phi)F_{\mu\nu}$ ensures the non-Newtonian nature within GR. It is known to source the precession of perihelion which is one among the three experimental tests of GR suggested by Einstein. Thus a non-Newtonian term, perceived with the Killing symmetries, needs further attention for its exploration. Though GR is a classical description of space-time curvatures, its inherent isometries elegantly add an intellectual dimension to a re-generated conserved quantity sourced by an appropriate coupling $(\phi F_{\mu\nu})$ of two distinct tensors. However its contribution turns out to be insignificant for a large scale structure of space-time but becomes significant for a small length scale phenomenon. It may entitle a non-Newtonian potential term to qualify for a quantum correction presumably in an underlying quantum theory of gravity. 

\sp
\noindent
A fact that a quantum correction is sensed via the isometries in GR is remarkable. It possesses a strength to explore new phenomena in GR and is believed to inspire an afresh perspective to a graviton. We explore one such possibility, leading to a gravitational wave/particle duality, among a few others in this paper. For instance the perihelion precession known in GR was re-visited by the authors in the recent past to ensure a non-perturbative (NP) correction to the (azimuthal) precession angle \cite{arXiv-NRS}. It was argued that an observed precession is essentially a non-planar effect and hence ensures more than one rotation. The non-commuting rotations imply a minimal length scale and hence its source is identified with the non-Newtonian term in $V_{\rm eff}$. A non-zero length scale allows a finite conjugate momentum and hence Heisenberg's uncertainty principle may be invoked for a non-Newtonian potential which in turn is believed to incorporate a quantum correction. It does not change the exact geometries in GR and hence turns out to be topological. The phenomenon further ensures a background independent correction and may lead to describe 
a NP gravity.

\sp
\noindent
 Interestingly a dipole, underlying the non-Newtonian term, ensures a NP correction to GR  \cite{arXiv-NRS}. Thus a NP gravity describes the motion of a dipole in a loop (say in $t$-channel) which in turn may viewed to 
govern the dynamics of a free graviton (in $s$-channel). Interestingly the perception for a graviton in a  NP gravity is apriori similar to that in 
an Open/Closed string duality \cite{Sen-sduality} but the underlying theories remain unparalled. In fact a graviton is a spin $2$ (mass-less) particle in a quantum theory of gravity whose low energy limit describes a metric field dynamics. Thus quantum gravity does not necessarily fix its underlying metric field description neither for a perturbative nor for a NP prescription. 

\sp
\noindent
Furthermore a non-Newtonian potential, $(\phi F_{\mu\nu})$ in an equatorial plane, underlying its topological nature, may take a form:  $(\phi$$ \wedge $$F_2)$. Intuitively it may incorporate Chern-Simon coupling $(A_1$$ \wedge $$F_2)$ in $d$$=$$3$ and a $(B_2$$ \wedge $$F_2)$ coupling in GR. The idea is in agreement with a fact that a mass-less (pseudo) scalar field dynamics is Poincare dual to that of a $2$-form in $d$$=$$4$. It provokes thought to believe for a $2$-form dynamics which in turn would like to replace the Newtonian scalar potential in GR. In fact the idea is well taken with a geometric torsion (GT) theory in $d$$=$$5$ bulk in the recent past \cite{JHEP-Abhishek}. Interestingly the bulk GT was shown to generate a boundary GR \cite{Physica-NDS}.  In this paper we briefly revisit and exploit a bulk GT for its possible quanta. We show that a bulk GT may govern a graviton in a boundary NP gravity.
In the context gauge theoretic tools have been explored in the last two decades to address the quantum gravity phenomena \cite{Desser-1982, wilczek1998, sugamoto2002-PTP,bern2010-PRL, Ko-Lin,  anastasiou2018-PRL}. Interestingly a dynamical generation of fourth or extra space dimension has been argued in GR \cite{arkani2001-PRL}.

\sp
\noindent
In this paper we address a pertinent issue sourced by the Killing symmetries to a gravitational potential. Our analysis reveals a quantum correction, realized via a topological coupling, to the GR. Presumably it is believed to describe the perihelion advances in GR. Interestingly we identify the topological term in the potential, with a dipole correction. Nonetheless the dipole is shown to be sourced by a conserved charge in a bulk (GT) on $R^{1,1}$$ \otimes $$S_3$. We compute the precession of a perihelion  with a renewed perspective in GT perturbation theory defined with an emergent metric \cite{JHEP-Abhishek} and estimate an extra space dimension. Results provide an evidence for a correspondence between a bulk GT  and a boundary GR phenomenon in presence of a dipole. In fact an afresh idea leading to a bulk ($2$-form) gauge theory and boundary (Einstein) gravity has already been discussed by two of the authors in a collaboration \cite{Physica-NDS}. Interestingly a propagating torsion underlying a modified gravity has recently been addressed \cite{nikiforova2018infrared}. Along the line a perihelion precession has  been revisited  for a plausible correction \cite{sultana2012,kupryaev2018concerning}.

\sp
\noindent
We plan the paper broadly in six sections. After a moderate introduction in section \ref{section01}, we briefly discuss the Killing symmetries leading to a dipole correction in section 2. We explore some of the essential features in a $2$-form gauge theory which in turn is shown to govern a bulk GT in section 3. A modified theory of gravity sourced by an axionic scalar in $d$$=$$5$ is discussed in sub-section 3.1. We provide a number of evidences leading to a bulk GT/boundary GR correspondence in a sub-section there. The Lorentz and Coulomb gauge conditions are exploited to argue for a plausible spin $2$ (mass-less) quantum in $d$$=$$5$ bulk GT. Furthermore the $2$-form theory is revisited with two fields, underlying in a braneworld scenario, to describe a boundary NP gravity. The bulk GT is investigated with $2$-form(s) ansatz on a braneworld which in turn ensures a black hole. In section 4, we perform a detailed analysis to compute the precession of perihelion in the bulk GT and show that the precession angle in GR receives a NP correction. The perspective of an EM field in the bulk GT is investigated in section 5 with an emphasis on the electric and magnetic components of a $4$-form. Interestingly the gravitational wave/particle duality is qualitatively analyzed to reveal a graviton in a NP gravity.
 
\section{Non-Newtonian potential: a key to NP gravity}
Killing symmetries, along with a constant of motion in GR, are briefly revisted to obtain effective potential $V_{\rm eff}$ on an equatorial plane. It is believe that $V_{\rm eff}$ may describe a generic gravitational potential. All terms in $V_{\rm eff}$ are checked for an aprior expectation for the Newtonian gravity on a plane within GR. It is indeed a pleasant surprise to notice that one term $V_q$ in $V_{\rm eff}$ turns out to be an exception to the Newtonian potential.

\sp
\noindent
In fact a non-Newtonian $V_q$ is believed to provide a clue to the quantum gravity. It is  primarily due to a fact that GR in a planar limit reduces to the Newtonian gravity. It is ensured by three conditions and they are: (i) linearized gravity, (ii) stationary and (iii) non-relativistic. Thus a non-planar effect incorporates the non-linearity and hence an extended (conserved) quantity is believed to source the GR. This in turn replaces a conserved (point) mass in Newtonian gravity to the energy in a relativistic formulation and hence ensures a deformation geometry $i.e.$ an arbitrary metric. Apriori an extended charge implies a minimal no-zero length scale and hence the Heisenberg's uncertainty principle may be invoked for a potential $V_q$ alone. However the presence of other terms ensuring Newtonian gravity in $V_{\rm eff}$ prohibits a quantum description and hence GR is indeed a classical (metric) field theory. Thus an essence of quantum gravity sourced by a non-Newtonian potential should be independent of the metric field dynamics leading to exact geometries in GR. It hints towards a NP gravity where $V_q$ is believed to describe a NP correction which is governed by the remaining terms in $V_{\rm eff}$. 

\subsection{Isometries and perihelion precession}
Consider a maximally symmetric vacuum solution in GR. In static coordinates the line element describes a Schwarzschild black hole: 
\be
ds^2=- \left (1-{{2{\tilde M}}\over{r}}\right ) dt^{2} +\left (1-{{2{\tilde M}}\over{r}}\right )^{-1}dr^{2}+r^{2}(d\theta^2+\sin^2\theta d\phi^2)\ ,\ \label{schwarzschild}
\ee
where ${\tilde M}$$=$$G_NM$ and $r$$>$$2{\tilde M}$. 
The geometry is characterized by one time-like Killing vector $K^{\mu}$$=$$(1,0,0,0)$ and three additional Killing vectors underlying the $S_2$-symmetry. The later describes the angular momentum vector and its magnitude is a conserved (Noether) charge $Q$. It takes a form $\xi^{\mu}$$=$$(0,0,0,1)$ and reflects a translational symmetry in $\phi$ while $K^{\mu}$ signifies a translational symmetry in $t$. Then the two conserved quantities are the energy E and the magnitude of angular momentum $Q$. They are given by
\be
E=-K_{\mu}{{dx^{\mu}}\over{{d\lambda}}}=\left (1-{{2{\tilde M}}\over{r}}\right ){{dt}\over{d\lambda}}\quad 
{\rm and}\quad Q=\xi_{\mu}{{dx^{\mu}}\over{d\lambda}}=r^2{{d\phi^2}\over{d\lambda}}\ .\label{conservedQ}
\ee
We begin with an equation for a constant of motion describing a time-like geodesics. It is given by
\be
g_{\mu\nu}{dx^{\mu}\over{d\lambda}}{dx^{\nu}\over{d\lambda}}=-1\ ,\qquad ({\rm where}\ \lambda={\rm affine\ parameter})\ .\label{motion-1}
\ee 
The Schwarzschild metric is used to re-express the equation in terms of the conserved quantities on an equatorial plane. It becomes
\be
\left (\frac{dr}{d\lambda}\right )^2 + \left (1-\frac{2G_NM}{r}\right )\left (1+\frac{G_NQ^2}{r^2}\right )= E^2\ .\label{Eq-1D}
\ee
Interestingly the equation may formally be identified to describe an one dimensional motion of  two unit of mass in an effective potential $V_{\rm eff}$ leading to a positive total energy $E^2$. In the case the $V_{\rm eff}$ on an equatorial plane takes a form:
\be
V_{\rm eff}=\left (1-\frac{2G_NM}{r}+\frac{G_NQ^2}{r^{2}}\right ) \ +\ \beta V_q\ ,\qquad {\rm where}\;\ V_q=-\frac{2G_N^2MQ^2}{r^3}\ .\label{motion-2}
\ee
The first and second terms in $V_{\rm eff}$ satisfy the inverse square law and hence correspond to the Newtonian gravitational potential. It is sourced by a scalar field $\phi({\mathbf{x}},t)$ and hence ensures a linear gravity. The third term may seem to be generated by a vector field $A_{\mu}({\mathbf{x}},t)$ which in turn may  sourced by a conserved (electric and/or magnetic) charge $Q$. The first three terms re-confirm the Newtonian gravity due to the gauge field equation of motion $\nabla^{\mu}F_{\mu\nu}$$=$$0$. Interestingly they may lead to define 
the gravitational potential $f(r)$ in a Reissner-Nordstr$\ddot{\rm o}$m (RN) black hole line element: $ds^2=-fdt^2+f^{-1}dr^2+r^2d\Omega^2$. It is important to notice that a constant of motion for a time-like geodesics modifies the causal sector in the line element without any change in 
the isometries. It may signify the self interaction in Einstein gravity which in turn is associated with the symmetric property of a metric field.  Analysis may suggest that the isometries may play an important role to re-define a vacuum in GR. 

\sp
\noindent
On the other hand the fourth term in $V_{\rm eff}$ defines a non-Newtonian potential as the  force does not satisfy the inverse-square  law. Thus the Killing symmetries allow a non-Newtonian gravity though the exact solutions do not. Generically a parameter $\beta$$=$$1$ describes GR while $\beta$$=$$0$ describes Newtonian gravity. Apriori a higher order in $G_N$ formally approves a consistency under a quantum correction due to a non-linear conserved quantity. 

\sp
\noindent
In the context $V_q$ is known to describe the observed perihelion precession of planet(s) in an approximately closed path around the Sun and generically for the precessing elliptical orbit around a star. It ensures that these  orbits are not perfect ellipses and hence they may allow a further possibility to explore the study of gravitational orbit from an alternate formulation underlying the perspectives in GR. In fact the precession of perihelion is one of the three experimental tests of GR suggested by Einstein. It may suggest that the observed precession of perihelion  advance possibly validates an alternate gauge theoretic formulation leading to a bulk GT. We postpone a perihelion precession analysis in a modified gravity underlying a bulk GT to a later section 4.

\sp
\noindent
We re-express the radial equation of motion (\ref{Eq-1D}) in terms of its variation in azimuthal angle. It is given by 
\be
\left (\frac{dr}{d\phi}\right )^2+(1-E^2)\frac{r^4}{G_NQ^2}+\left (1-\frac{2Mr}{Q^2}\right )r^2-2G_NM=0\ .\label{Eq-1D-2}
\ee
It may be re-expressed as a second order  inhomogeneous differential equation in two steps: (i) with a change of variable $X=Q^2(Mr)^{-1}$, which is 
possibly due to a lower bound on $r$ ensured by the $V_q$ term, followed by (ii) a differentiation w.r.t. the azimuthal angle $\phi$. It becomes
\be
{{d^2X}\over{d\phi^2}} - \left ({{3G_NM^2}\over{Q^2}}\right )X^2 + X=1\ .\label{Eq-1D-3} 
\ee
A generic solution may be approximated with $X=X_N + Y$, where $X_N$ signifies the differential equation for Newtonian gravity only. The $Y$ is an infinitesimally small deviation to $X_N$ in GR. Then the equation is splitted into two independent and are given by
\be
{{d^2X_N}\over{d\phi^2}}+X_N=1\qquad {\rm and}\quad {{d^2Y}\over{d\phi^2}}+Y= \left ({{3G_NM^2}\over{Q^2}}\right )X_N^2\ .\label{Eq-1D-4}
\ee
The solution to Newtonian gravity is worked out to yield: $X_N$$=$$(1$$+$$e \cos\phi)$, where $e$= eccentricity of an ellipse. It is used for GR in the second equation to obtain a solution \cite{carroll} for $Y$. A significant term in $Y$ is used to obtain an approximate solution:
\be
X=1+ e\cos[(1-\delta)\phi]\ ,\qquad {\rm where}\quad \delta= \left ({{3G_NM^2}\over{Q^2}}\right )\ .\label{Eq-1D-5}
\ee
Thus $\delta$ ensures an advancement $\triangle \phi$ in azimuthal precession angle during each orbit of a planet around the Sun. It is given by
\be
\triangle\phi= 2\pi\delta= 6\pi G_N\left [{{M}\over{Q}}\right ]^2\ .\label{Eq-1D-5} 
\ee
Thus a ratio, of the conserved charges, determines an order scale of a precession angle for a perihelion. An angular velocity of a planet along an orbit assigns a non-zero $Q$. A larger angular momentum ensures a smaller precession angle. In a special case for an extremal geometry underlying a geodesic, the precession angle takes a minimal value $(6\pi G_N)$.
\begin{figure}[h]
\centering
\mbox{\includegraphics[width=0.7\linewidth,height=0.2\textheight]{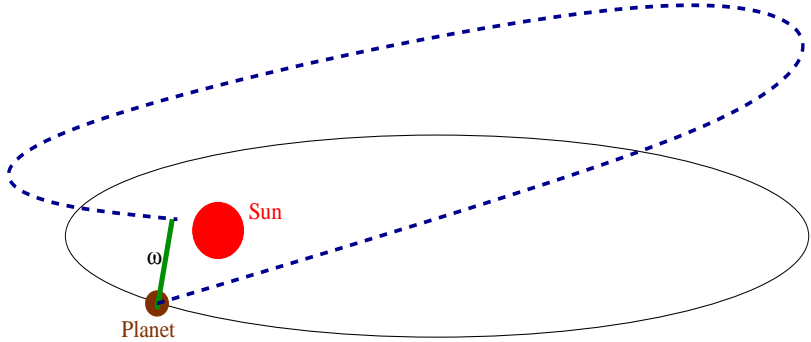}}
\caption{\it A schematic diagram shows perihelion precession in GR for a planet around the Sun}
\end{figure}

\subsection{Dipole correction to GR}
Interestingly the non-Newtonian potential $V_q$  may be re-interpreted in terms of a mass dipole $(M_D)$ defined with a dipole moment $D$$=$$MQ$. It is given by
\be
V_q= QM_D\qquad {\rm and}\quad M_D= G_N {{2D}\over{r^3}}\ .\label{M-dipole}
\ee
The Newton's coupling identifies the mass dipole and $M$$ \neq $$0$ ensures a dipole moment $D$.
The mass dipole is sourced by the conservation of both energy $E$ and charge $Q$ but turns out to be insignificant for large $r$ geometries. However for small $r$, the $V_q$ can incorporate a vital (relativistic) correction to the geometry underlying a classical vacuum in GR. 

\sp
\noindent
Thus a mass dipole, underlying a non-Newtonian potential, consistently incorporate a NP quantum gravity phenomenon. In principle a non-Newtonian potential can describe an interacting non-point masses and hence incorporates a lower cut-off on the radial distance $r$.  A dipole ensures the background independence of the potential $V_q$ on an equatorial plane and  may be interpreted as a NP correction.

\sp
\noindent
 Analysis shows that the Killing symmetries  may provide a remarkable clue to unfold a topological correction atleast to the maximally symmetric family of black holes defined with more than one tensor field source. For instance, the second and third terms in $V_{\rm eff}$ ensure the Reissner-Nordstr$\ddot{\rm o}$m (RN) black hole. The idea may lead to believe for a topological correction to the exact solution(s) in GR. It may inspire one to formally propose an action to describe a NP-gravity. For a coupling $l$$=$$[{\rm length}]$, it may be given by
\be
S={1\over{4}}\int d^4x{\sqrt{-g}}\left ({{\cal R}\over{4\pi G_N}} - F_{\mu\nu}^2 - {1\over{3l^2}}H_{\mu\nu\lambda}^2\right )\ 
 -{1\over{4\pi G_N}}\int B_2\wedge F_2\ ,\label{q-action}
\ee
\be
{\rm where}\quad F_{\mu\nu}=(\nabla_{\mu}A_{\nu}-\nabla_{\nu}A_{\mu})\qquad {\rm and}\quad H_{\mu\nu\lambda}=(\nabla_{\mu}B_{\nu\lambda} + {\rm cyclic})\ . \nonumber
\ee
The $BF$-term incorporates a topological coupling to the metric dynamics via the gauge fields. For a detailed study on $BF$-gravity see refs\cite{krasnov2009,mielke2010-PLB, Celada2016}. A consistent truncation of the generic action (\ref{q-action}) yields a topologically coupled Einstein-Maxwell theory.

\sp
\noindent
 Interestingly the dynamical terms in the action may alternately be derived from the $d$$=$$5$ Einstein gravity on $S^1$. It re-ensures that a higher dimensional gravity can be a potential tool to address a quantum gravity phenomenon in a lower dimension. However the $BF$-coupling in the action (\ref{q-action}) underlying the quantum correction needs to be placed by hand as it cannot be derived from $d$$=$$5$ (Kaluza-Klein) gravity. 

\section{Geometric Torsion in $d\ge 5$ bulk}
\subsection{Modified gravity}
In principle the commutator of derivatives in a theory is known to describe the (gauge or gravity) curvatures. For instance a non-commutative (NC)  space-time $[x^{\mu},x^{\nu}]=(4i)\theta^{\mu\nu}$ ensures that its canonical conjugate momenta satisfy $[p_{\mu},p_{\nu}]\neq 0$ and hence $[\partial_{\mu},\partial_{\nu}]\neq 0$. It is known to describe a new geometry with a non-vanishing commutator: 
\be
\left [A_{\mu}\ ,A_{\nu}\right ]=-i\left (F_{\mu\alpha}\theta^{\alpha\beta}F_{\beta\nu}\right )\ .\label{NC-1}
\ee
Thus self-interactions are elegantly incorporated into an $U(1)$ theory and the manifest gauge invariance appears to be broken! However the $U(1)$ invariance is beautifully perceived with a modified gauge transformation \cite{seiberg-witten} and hence the NC gauge theory describes a new notion of curvature. 

\sp
\noindent
In the context GR is a classical metric $g({\mathbf{x}},t)$ field theory. It is intrinsically defined with the Christoffel connections:
\be
{\Gamma}_{\mu\nu}^{\lambda}={1\over2}g^{\lambda\rho}(\partial_{\mu}g_{\rho\nu}+\partial_{\nu}g_{\mu\rho}-\partial_{\rho}g_{\mu\nu})\ .
\ee 
The connections modify $\partial_{\mu}\rightarrow\nabla_{\mu}$. The commutator of the covariant derivatives acting respectively on a scalar field $\Phi({\mathbf{x}},t)$ and on a vector field $A_{\mu}({\mathbf{x}},t)$ may be worked out to yield:
\be
[\nabla_{\mu}\ ,\nabla_{\nu}]\Phi = 0 \qquad {\rm and}\quad [\nabla_{\mu}\ ,\nabla_{\nu}]A_{\lambda} ={\cal R}_{\mu\nu\lambda}{}^{\rho}A_{\rho}\ ,\label{commutator} 
\ee
where the Riemann curvature tensor: 
\be 
{\cal R}_{\mu\nu\lambda}{}^{\rho} = (\partial_{\nu}\Gamma_{\mu\lambda}^{\rho}-\partial_{\mu}\Gamma_{\nu\lambda}^{\rho} +\Gamma_{\nu\sigma}^{\rho}\Gamma_{\mu\lambda}^{\sigma}-\Gamma_{\mu\sigma}^{\rho}\Gamma_{\nu\lambda}^{\sigma})\ .
\ee
We would like to re-emphasize a few interesting points underlying an elegant geometric formulation by Einstein. The first and the second terms in ${\cal R}_{\mu\nu\lambda}{}^{\rho}$ ensure the dynamics of ${\Gamma}_{\mu\nu}^{\lambda}$ field presumably in a $\partial_{\mu}$-description which in turn describes the dynamics of a metric field in a second order. The third and fourth terms signify the metric field dynamics in the same description. However they cannot be treated independently as each of them break the tensor transformation property. Needless to mention that together they retain the tensor behaviour of ${\cal R}_{\mu\nu\lambda}{}^{\rho}$.  Secondly a vanishing commutator in (\ref{commutator}) ensures that a scalar field theory alone in $\nabla_{\mu}$-description does not fetch the dynamical aspects of metric field in GR. The $\Phi$ (a rank zero tensor) field being linear, it does not play a significant role in a nonlinear theory. A  non-vanishing commutator unfolds a fact that $A_{\mu}$ can alternately be realized as a non-linear gauge field. The intriguing observation in Einstein's gravity is consistent with the perceived new geometry underlying a NC gauge theory \cite{seiberg-witten}. Generically it makes a non-zero rank tensor  special and provokes thought to believe in the success of a gauge theory to presumably describe the Einstein gravity phenomenon in an alternate prescription. Interestingly a $2$-form underlying (pseudo) scalar in $d$$=$$4$ is in  equal footing to that of a scalar field sourcing a Newtonian potential in GR.

\sp
\noindent
Along the line of thought, a geometric torsion curvature ${\cal K}_{\mu\nu\lambda}{}^{\rho}$ has been worked out underlying a $2$-form dynamics in the recent past by one of the author in a collaboration \cite{JHEP-Abhishek}. In particular a dynamical $2$-form $B_{\mu\nu}({\mathbf{x}},t)$ leading to a field strength  $H_{\mu\nu\lambda}$$=$$3\nabla_{[\, \mu}B_{\nu\lambda ]}$ in an $U(1)$ gauge theory has been treated as a torsion connection to construct a modified covariant derivative ${\cal D}_{\mu}$. Generically a modification to  GR, leading to a geometric torsion ${\cal H}_{\mu\nu\lambda}$$=$$3{\cal D}_{[\, \mu}B_{\nu\lambda ]}$, has been proposed in a higher dimension $d$$ \ge $$5$ for a non-perturbative ${\cal D}_{\mu}$. It is straight-forward to check that the ${\cal H}_{\mu\nu\lambda}$ retains the $U(1)$ gauge invariance under the usual transformation of  a $2$-form in a NP-description. In particular the bulk GT dynamics underlying the ${\cal K}_{\mu\nu\lambda}{}^{\rho}$ was worked out in a minimal $d$$=$$(4+1)$. Interestingly the bulk GT was shown to describe some of the GR phenomena \cite{PRD-Abhishek}. Explicitly 
\bea
2{\cal D}_{\mu} B_{\nu\lambda}&=&2\nabla_{\mu}B_{\nu\lambda}+H_{\mu\nu}{}^{\rho}B_{\rho\lambda} - H_{\mu\lambda}{}^{\rho}B_{\rho\nu}\nonumber\\
{\cal H}_{\mu\nu\lambda}&=&H_{\mu\nu\lambda}+3H_{[\mu\nu}{}^{\rho}B_{\rho\lambda ]}\ .\label{modifiedD}
\eea
Generically a GT modifies the gauge theoretic torsion due to the topological coupling terms. Thus a GT essentially governs a topologically massive $2$-form gauge theory. Interestingly the Lorentz scalar $({\cal H}_{\mu\nu\lambda}{\cal H}^{\mu\nu\lambda})$ generates a mass term for the $2$-form in the Lagrangian density \cite{PTEP-NS} and hence is believed to describe a topologically massive perturbation theory. The commutators in eq(\ref{commutator}) under $\nabla_{\mu}\rightarrow {\cal D}_{\mu}$ has been shown to incorporate new curvatures \cite{JHEP-Abhishek}. They are worked out in presence of a coupling, $i.e.$ for $H_3\rightarrow {\cal H}_3$, to yield:
\be
[{\cal D}_{\mu}\ ,{\cal D}_{\nu}]\Phi = {\cal H}_{\mu\nu}{}^{\rho}{\nabla}_{\rho}\Phi \qquad {\rm and} \quad [{\cal D}_{\mu}\ ,{\cal D}_{\nu}]A_{\lambda}=
\left ({\cal R}_{\mu\nu\lambda}{}^{\rho} + {\cal K}_{\mu\nu\lambda}{}^{\rho}\right )A_{\rho}\ ,\label{curvature-1}
\ee
\be
{\rm where}\qquad {\cal K}_{\mu\nu\lambda}{}^{\rho}= {1\over4}\left ({\cal H}_{\mu\lambda}{}^{\sigma}{\cal H}_{\nu\sigma}{}^{\rho}
-{\cal H}_{\nu\lambda}{}^{\sigma}{\cal H}_{\mu\sigma}{}^{\rho}\right )\ + {1\over2}\left (-\nabla_{\nu} 
{\cal H}_{\mu\lambda}{}^{\rho}+\nabla_{\mu}{\cal H}_{\nu\lambda}{}^{\rho}\right )\ \label{curvature-2}
\ee
Firstly, a NP scenario is defined with ${\cal D}_{\mu}$ only. 
It evolves with a generic fourth rank curvature tensor in addition to the Riemannian tensor. Thus ${\cal K}_{\mu\nu\lambda}{}^{\rho}$ incorporates a NP correction to the Einstein-Hilbert action in particular and to a number of theories defined with various irreducible curvature tensors, and/or their appropriate combinations, derived from the Riemannian tensor. Generically the new curvature tensor modifies the geometric formulation of gravity and hence is identified with a modified theory of gravity. A modification to Einstein gravity was constructed by one of the author in a collaboration \cite{JHEP-Abhishek}. It was shown that the torsion connection consistently modifies the covariant derivative $\nabla_{\mu}$ in Einstein gravity to ${\cal D}_{\mu}$ which satisfies ${\cal D}_{\mu}g_{\nu\lambda}$$=$$0$.

\sp
\noindent
Secondly, a commutator in  eq(\ref{curvature-1}) may ensure a (scalar field) dynamical correction to the Einstein gravity. In a special case for a condensate $<$$ \Phi $$>$$ \neq $$0$ the dynamical correction vanishes to yield the Einstein gravity. Interestingly the scenario leading to a GT in $d$$=$$5$ has been worked out in the recent past \cite{JHEP-Abhishek}. Though the $U(1)$ gauge invariance is spontaneously broken by an axionic scalar condensate,  it has been shown to be restored with an emergent metric $(g_{\mu\nu}-l^2{\cal H}_{\mu\alpha\beta}{\cal H}^{\alpha\beta}{}_{\nu})$ in a $\nabla_{\mu}$-perturbation theory. The length $l$ signifies a minimal scale $<$$ \Phi $$>$. It is believed to exclude the Newtonian gravity and hence can be a plausible candidate to describe a quantum gravity phenomenon. Nevertheless a bulk condensate decouples from the boundary under a  bulk GT/boundary GR correspondence \cite{Physica-NDS}. 

\subsection{Evidence for Bulk GT/Boundary GR}
Interestingly, the first and second terms (say $K_{\mu\nu\lambda}{}^{\rho}$) in the generic curvature (\ref{curvature-2}) precisely share the anti-symmetric (within first and second pairs) and pair-symmetric properties of the Riemann tensor. It ensures a minimal space-time dimension $d$$=$$5$ to a GT in bulk. Interestingly the $4$-form coupled to Einstein gravity is known for a dynamical generation of the cosmological constant $\Lambda$ on $S_1$ \cite{dynConst}. It may turn out instrumental to unfold an origin of dark energy in the universe. 

\sp
\noindent
A generic correspondence between a bulk GT and a boundary (Einstein) gravity phenomenon is based on a number of evidences. For instance  non-Riemannian space-time curvature tensor $K_{\mu\nu\lambda\rho}$, in absence of a propagating torsion in bulk, has been shown to share all the properties of the Riemannian $R_{\mu\nu\lambda\rho}$ under the interchange of its indices \cite{JHEP-Abhishek}. Recall that the space-time curvature is an observable and the potentials (metric and $2$-form) are not. Thus an observer would not be abled to distinguish between the Einstein gravity and its alternate formulation with $2$-form(s) gauge theory. In fact the scalar curvature $K$ computed from a $2$-form ansatz in the bulk GT  identifies with an expression for the scalar $(C_{\mu\nu\lambda\rho}C^{\mu\nu\lambda\rho})$ in GR for a static vacuum, where $C_{\mu\nu\lambda\rho}$= conformal-weyl tensor. Generically the local degrees of a mass-less $2$-form on $R^1\otimes S_{d}$ is precisely equal to that of a metric on $S_{d}$ where the radius of $S_d$ may be identified with a pseudo-scalar field $\chi$. 

\subsubsection{Torsion curvatures}
The irreducible curvatures have been worked out from the reducible tensor $K_{\mu\nu\lambda}{}^{\rho}$. They may seem to govern  two tensors $K_{\mu\nu}$ and $K$. They are:
\be
K_{\mu\nu}=-{1\over4} {\cal H}_{\mu\alpha\beta}{\cal H}^{\alpha\beta}{}_{\nu}\qquad {\rm and}\qquad K=-{1\over4} {\cal H}_{\mu\nu\lambda}{\cal H}^{\mu\nu\lambda}\ .\label{curvature-3}
\ee
The third and fourth terms in eq(\ref{curvature-2}) ensure a propagating GT underlying a non-trivial ${\cal F}_4$$=$$d{\cal H}_3$ in a perturbative prescription \cite{PTEP-NS}. In particular they define curvatures: 
\be
L_{\mu\nu\lambda\rho}=\frac{1}{2} {\cal F}_{\mu\nu\lambda\rho} 
+ \frac{1}{2}\left (\nabla_{\rho}{\cal H}_{\mu\nu\lambda} - \nabla_{\lambda}{\cal H}_{\rho\mu\nu}\right )\quad {\rm and}\;\ L_{\mu\nu}= {\rm Tr} \left (L_{\mu\nu\lambda\rho}\right )= -{1\over{2}}\nabla^{\lambda}{\cal H}_{\lambda\mu\nu}\ .\label{Beqn}
\ee
An on-shell $B_{\mu\nu}$ implies $L_{\mu\nu}$$=$$0$ and hence $K_{\mu\nu\lambda}{}^{\rho}\rightarrow {\cal F}_{\mu\nu\lambda}{}^{\rho}$. Thus 
${\cal K}_{\mu\nu\lambda}{}^{\rho}$ may equivalently be represented by $[K, K_{\mu\nu}, {\cal F}_{\mu\nu\lambda}{}^{\rho}]$. Among a number of choices for modification to the Riemannian geometry, a modified Einstein gravity has been argued with $[{\cal R}, K, {\cal F}_4]$ curvatures in a NP gravity. In a torsion decoupling limit, the NP description reduces to a perturbation $[{\cal R}, F_4]$ theory. Remarkably the curvatures in a decoupling (or low energy) limit are in agreement with the bosonic part of $d$$=$$11$ supergravity \cite{cremmer}. An aprior analysis reveals that the bosonic field contents in GT may lead to a non-perturbation M-theory in $d$$=$$11$ whose low energy limit is known to describe the $N$$=$$1$ supergravity.

\sp
\noindent
Under a special case for a non-propagating GT  the $4$-form vanishes. In the limit, a NP prescription evolves with the curvatures $K_{\mu\nu}$ and $K$ only. They may appear indistinguishable from the Ricci tensors ${\cal R}_{\mu\nu}$ and ${\cal R}$ respectively. This is due to a fact that the curvatures in both gauge and Einstein gravity are observable but their respective tensor potentials $2$-form and metric are not. It may provoke thought to believe for an alternate prescription underlying a bulk GT describing the observed phenomena in GR. In particular the equivalence between the $d$$=$$4$ non-Riemannian and $d$$=$$3$ Riemannian (meaning only the Ricci) curvatures turns out to be precise. We recall that a derived non-Riemannian curvature tensor in $d$$=$$4$ freezes the propagation of a GT and hence is a special case. 

\sp
\noindent
Generically a bulk GT theory is defined with $d$$ \ge $$5$ as a local degree for GT re-ensures a minimal $d$$=$$5$. This is similar to a fact that a local degree for a metric field ensures a minimal $d$$=$$4$ and hence the GR. Analysis may formally identify a fundamental role of a $4$-form in bulk GT to that of a conformal-Weyl tensor $C_{\mu\nu\lambda\rho}$ in GR. Interestingly  a $2$-form leading to an emergent Schwarzschild geometry, has been shown to describe a scalar  curvature $K$$  \propto  $$r^{-6}$.  This is  agreement with the  curvature $({\cal R}_{\mu\nu\lambda\rho}{\cal R}^{\mu\nu\lambda\rho})$ for an identical geometry in GR \cite{PRD-Abhishek}.

\subsubsection{An apparent $2$-form $\rightarrow$ a ``spin $2$'' particle}
A count for the propagating degrees of a mass-less $2$-form in the bulk $(d$$+$$1)$ precisely matches with that of a metric field (coupled to a scalar field) dynamics in $d$-dimensions. It may suggest that a scalar-(metric)tensor theory in $d$-dimensions may equivalently be described by a $2$-form $U(1)$ theory in $(d$$+$$1)$. The later may also be viewed as a massive $2$-form in $d$-dimension. Intuitively it may prompt one to identify a graviton (mass-less spin $2$) with the quantum of a massive $2$-form! In fact a $2$-form equation of motion $\nabla^{\lambda}H_{\lambda\mu\nu}$$=$$0$ in an $U(1)$ gauge theory defined with a background gravity  may be re-expressed as:
\be
\nabla^2B_{\mu\nu} + F_{\mu\nu}=0\ ,\qquad {\rm where}\;\; F_{\mu\nu}=\left (\nabla_{\mu}C_{\nu}-\nabla_{\nu}C_{\mu}\right ) .\label{B-1}
\ee
An identification with an one form incorporates $d$-number of constraints in a $2$-form gauge theory. Apriori the local degrees for a $2$-form becomes $D_2$$=$$[d(d$$-$$3)/2]$. However the Lorentz condition $\nabla^{\mu}C_{\mu}$$=$$0$, under an identification $C_{\mu}$$=$$\nabla^{\alpha}B_{\alpha\mu}$, is consistently enforced by the background gravity $i.e.\ [\nabla^{\mu},\nabla^{\nu}]B_{\mu\nu}$$=$$0$ and hence the counting of local degrees become subtle. 
On the one hand the Lorentz condition ensures a propagating Goldstone scalar $\nabla^2\Phi $$=$$0$ and hence the total number of local degrees for a massive $2$-form turns out to be $[(d$$-$$1)(d$$-$$2)/2]$. 

\sp
\noindent
Interestingly a background gravity ensures the Lorentz gauge condition automatically. This in turn may allow an alarming possibility of $D_2$-number of local degrees for a massive $2$-form! Interestingly $D_2$ equals to that of a metric field which is believed to describe a graviton. We recall that: (i) a photon (mass-less spin one and hence $2$-polarizations) and a massive spin one vector boson ($3$ spin-polarizations) are respectively governed by two and three local degrees of $C_{\mu}$ in $d$$=$$4$ and similarly (ii) a graviton (mass-less spin $2$ and hence $2$-polarizations) is governed by two local degrees of $g_{\mu\nu}$ in $d$$=$$4$.

\sp
\noindent
 In the context the equations of motion (\ref{B-1}) when $C_{\mu}$$=$$\nabla_{\mu}\Phi,\ i.e.$ in a pure gauge, assigns $D_2$$=$$5$ local degrees to an apparent $2$-form in the bulk GT for a minimal dimension $d$$=$$5$. The local degree of the Goldstone scalar $\Phi$ is believed to be absorbed by a  mass-less $2$-form (Poincar${\acute{\rm e}}$ dual to $1$-form) to describe a massive $2$-form with $6$ local degrees in $d$$=$$5$. An analogy drawn along the line of the quanta (photon and graviton) may provoke thought to believe that an apparent $2$-form can be a potential candidate to govern a spin $2$ (mass-less) quantum! This observation is in agreement with the bulk ($2$-form)/boundary gravity correspondence \cite{Physica-NDS}. Interestingly one local degree by the Goldstone scalar $\Phi$ is believed to ensure a transverse dimension to the boundary gravity. For a condensate $<$$ \Phi $$>$, the bulk torsion decouples and the boundary NP gravity reduces to GR. Interestingly a bulk GT perspective to a boundary GR naturally assigns a transverse nature to the gravitational wave which is otherwise an assumed phenomenon in a linearized GR.

\sp
\noindent
Recall that a space-time covariance is broken by hand in an $U(1)$ gauge theory when it describes an electro-magnetic (EM) field. Thus a Lorentz condition is splitted to yield Coulomb conditions: $C_{0}$$=$$0$ and $\nabla^{i}C_{i}$$=$$0$. They are indeed the $2$-form gauge conditions:
\be
B_{i0}=0\qquad {\rm and}\qquad \nabla^iB_{ij}=0\ .\label{Coulomb}
\ee 
Aprior eight constraints in $d$$=$$5$ become actually be seven due to an over counting by the second expression in eq(\ref{Coulomb}). In the case a $2$-form is thus governed by three local degrees. As discussed, a propagating Goldstone scalar arised out of the gauge symmetries  add a local degree to the mass-less $2$-form which is Poincar$\acute{\rm e}$ dual to a gauge field $A_{\mu}$. This in turn describes a massive $A_{\mu}$ theory and spontanesouly breaks the duality symmetry between a $2$-form and the $A_{\mu}$ gauge theories. 

\subsubsection{Composite particle with an axion}
On the other hand a propagating GT in bulk may alternately be viewed to govern a pseudo scalar field $\chi$ via Poincar$\acute{\rm e}$ duality:
\be
\nabla^{\alpha}\chi= {1\over{24{\sqrt{-g}}}}\epsilon^{\alpha\mu\nu\lambda\rho}{\cal F}_{\mu\nu\lambda\rho}\ .\label{Pduality}
\ee
The QFT of $\chi$ describes a pseudo particle called axion. A change in space-time signature, enforced by the duality in odd dimensions, apriori ensures an axionic ghost  (a phantom) in the bulk $2$-form gauge theory! Nevertheless a gauge field can as well change the space-time signature which is known to replace the original signature equivalently. Thus an axion in a bulk gauge theory becomes physical though it can govern a phantom in $d$$=$$5$ Einstein gravity. This inspiring fact allows one to uplift the $d$$=$$4$ equivalence (in local degrees) between the GR and a mass-less  
$(A_{\mu})$ gauge theory to $d$$=$$5$ Einstein gravity (coupled to a $4$-form field strength) and a massive $A_{\mu}$ theory. Interestingly an effective local degree of a metric field is always reduced by a phantom and in the case it turns out to be four. Alternately in a similar way an equivalence may be established in $d$$=$$4$ between a topologically massive $A_{\mu}$ theory and the GR. A topologically mass term does not modify the equation of motion of the gauge field but is empowered to modify a global property of a conserved charge sourcing the gauge field. In the case the bulk dynamics of an axion ensures a boundary topological correction $({\cal H}_3$$ \wedge $$d\chi)$ to  GR underlying a proposed bulk GT/boundary GR correspondence. Generically a  topological term incorporates a NP correction to the GR and together they may be re-interpreted as a boundary NP  gravity. 

\sp
\noindent
Interestingly a $2$-form in the Coulomb gauge along with a complex scalar field (formed from a Goldstone scalar and an axionic scalar) may be proposed to govern a mass-less ``composite'' particle of spin $2$ in the bulk GT. Apriori the proposal for a composite particle is in agreement with an apparent $2$-form in Lorentz gauge which in turn re-ensures a gauge independent approach to obtain a graviton under a bulk GT/boundary NP gravity phenomenon. 
One may believe that a composite particle presumably shares a spin $2$ characteristic of a graviton in $d$$=$$5$ bulk. Intuitively a (Goldstone) scalar particle (spin zero) underlying an oscillating circular profile of waves superimpose with an electric field oscillations to produce a group of waves oscillating formally on ${\tilde S}_1$$ \otimes $${R}$, where ${\tilde S}_1$  denotes  varying area under $S_1$. Similarly a superposition of axionic scalar waves on an oscillating pseudo vector (magnetic field ${\mathbf{M}}$) lead to a wave profile on ${\tilde S}_1$$ \otimes $${R}$. It is perpendicular to the group profile along ${\mathbf{E}}$. These two polarizations, underlying a composite field, lead to a transverse  wave for a group covered with a circular profile which in turn reduces the periodicity to half of that of an EM wave. An empirical formula with a reduced periodicity assigns a spin $2$ to its dual in the boundary NP. Interestingly an apparent $2$-form discussed  formally with Lorentz condition in section 2.2.2 shares the spin $2$ property of a composite field with Coulomb gauge conditions. It further provides an evidence to a proposed bulk GT/boundary GR.

\sp
\noindent
In the context we recall the ${\rm AdS}_5/{\rm CFT}_4$ duality in superstring theory \cite{susskind-hologram, witten-AdS, maldacena-AdS}. It is known to correspond a weakly coupled gravity in bulk to a strongly coupled (super-symmetric) gauge theory on boundary. A strong-weak coupling duality envisaged an equivalence between a perturbative theory in bulk and a boundary NP theory. Remarkably this underlying essence of perturbation bulk/boundary NP correspondence has also been carried forward in a proposed duality between a $2$-form gauge theory in $d$$=$$6$ bulk and a boundary (Einstein) gravity \cite{Physica-NDS}. A traceless energy-momentum-stress tensor for a $2$-form ensures a conformal symmetry (${\rm CFT}_6$) atleast in the classical theory. Generically the  conjectured duality is believed to identify a bulk GT or a $2$-form gauge theory on a specified topology $R^1$$ \otimes $$S_d$. A bulk on a tensor product space of a causal and a maximally symmetric space defines a boundary on $S_d$. Since the Killing vector under a time translation formally resembles to that of the azimuthal coordinate $\phi$, the Wick rotation $\phi $$ \rightarrow $$it$ recreats a real time on the boundary and hence $S_d$ apriori maps  to $R^{0,1}$$ \otimes $${\tilde S}_d$, where ${\tilde S}$ signifies an half spatial section or a semi-spherical symmetry under a newly identified azimuthal angle $\psi$ but for  $0$$<$$\psi$$<$$\pi$. Intuitively a semi-spherical space ${\tilde S}_d$ under a vacuum or stable nucleation reshapes to (a local) $S_{(d-1)}$. 

\sp
\noindent
Remarkably the holographic idea (bulk GT/Boundary gravity) does not necessarily restrict the boundary geometry to an ${\rm AdS}$ rather it allows all (positive, negative and vanishing) values of $\Lambda$. Thus the success of gauge theoretic tools in bulk would immensely be helpful to explore the Friedmann-Lema$\hat{\rm i}$tre-Robertson-Walker (FLRW) universe under the higher form(s) dominance. As a bonus the boundary gravity is not bounded, $i.e.$ may not be an isolated system globally, which ensures its interacting nature realized via Newtonian gravity. Nevertheless a local boundary leading to an ${\rm AdS}$ patch within may be nucleated in a NP gravity \cite{JHEP-Abhishek}. Arguably a $2$-form or even a higher form quantum in an $U(1)$ gauge theory, in presence of a background black hole, is believed to create a vacuum pair of Universe/anti-Universe $(U{\bar U})$ across the event horizon following the principle of Schwinger mechanism \cite{schwinger}. Recall a fact that the pair creation is a NP mechanism. The momentum conservation ensures that an universe is moving away from the anti-universe along an hidden transverse space dimension. The repelling $U$ and ${\bar U}$ exchange (quantum) information via an axionic scalar dynamics in a NP scenario. For instance an axionic condensate fixes the radius of a higher dimensional sphere which leads to a decoupling of $U$ from ${\bar U}$ or a decoupling of the topological correction. Alternately a torsion decoupling limit in a NP-gravity re-confirms that GR can be a boundary phenomenon.

\sp
\noindent
Very recently a perihelion precession was computed in $d$$=$$5$ bulk GT formulation \cite{arXiv-NRS}. Arguably the advances in azimuthal precession angle is perceived well along an elliptically elongated spiral path in a bulk GT. The precession has been shown to retain its form in GR but modifies non-perturbatively.  Our analysis leading to a topological correction to the computed precession in GR may validate the conjectured bulk GT/boundary NP gravity \cite{Physica-NDS}. It further ensures a consistent description to the gravitational wave/particle duality.

\subsection{Braneworld scenario: a gravitational pair}
The mentioned difficulty may be resolved with a proposed correspondence \cite{Physica-NDS} between a $2$ form perturbation theory, underlying a conformal symmetry, in $d$$=$$6$ bulk and a boundary ${\rm AdS}_5$. The idea has been modelled in a $U(1)$ gauge theory described by a $2$-form in presence of a background gravity. On $S_1$ there are two massless $2$-forms and the gauge group becomes $U(1)$$ \otimes $$U(1)$ in the $d$$=$$5$ bulk. Now the $2$-forms in the bulk on $R^{(1,1)}$$ \otimes $$S_3$ under the bulk GT/boundary GR correspondence 
is described with all terms in $V_{\rm eff}$ in eq(\ref{motion-2}). 

\subsubsection{Nonperturbation gravity}
We consider a scenario described by two different $2$-forms $(B_{\mu\nu}$ and ${\cal B}_{\mu\nu})$ respectively defined with the covariant derivatives $\nabla_{\mu}$ and ${\cal D}_{\mu}$ in a $(4$$+$$n)$-dimensional gauge theory. We consider $\nabla_{\lambda}{\cal B}_{\mu\nu}$$=$$0$ and ${\cal D}_{\lambda}B_{\mu\nu}$$=$$0$, $i.e.$ a covariantly constant ${\cal B}_{\mu\nu}$ with $\nabla_{\mu}$ description where $H_3$$=$$dB_2$ is a gauge theoretic torsion. A covariantly constant $B_{\mu\nu}$ in ${\cal D}_{\mu}$ description defines a geometric torsion ${\cal H}_3$$=$$d^{\cal D}{\cal B}_2$. Thus in case of a pure ${\cal D}_{\mu}$ derivative theory, $B_{\mu\nu}$ behaves as a background field (meaning non-dynamical) and similarly in an original theory, ${\cal B}_{\mu\nu}$ turns out to be a background. However both the $2$-forms are dynamical in the bulk action (\ref{torsion-bulk}). In addition a non-zero ${\cal F}_4$$=$$d{\cal H}_3$ together with both the $2$-forms describe a scenario leading to an action:
\be
S={{-1}\over{48\lambda^2}}\int_{\Sigma}d^{(4+n)}y{\sqrt{-g}}\ \Big (l^2{\cal F}_{\mu\nu\lambda\rho}^2 +4 {\cal H}_{\mu\nu\lambda}^2 + 4 H_{\mu\nu\lambda}^2\Big )\ , 
\label{torsion-bulk}
\ee
where ${\lambda}^2$$=$$l^{(n+2)}$. In the case the $B_{\mu\nu}$ behaves as a background in a ${\cal D}_{\mu}$ derivative theory. Similarly in an original theory the ${\cal B}_{\mu\nu}$ turns out to be a background. They respectively lead to a NP and perturbation description. In fact they have been imagined to govern a brane world underlying two independent $2$-forms for an axionic condensate \cite{JHEP-Abhishek}. The energy-momentum-stress (EMS) tensor $T_{\mu\nu}=-4\delta S/({\sqrt{-g}}\delta g_{\mu\nu})$ in the perturbation theory becomes: 
\begin{eqnarray}
T_{\mu\nu}&=&{{l^2}\over{3}}\left ( {\cal F}_{\mu\alpha\lambda\rho}{{\cal F}_{\nu}}^{\alpha\lambda\rho}-
{{g_{\mu\nu}}\over{8}} {\cal F}_{\alpha\beta\lambda\rho}^2\right )\nonumber \\
&+& \left ({\cal H}_{\mu\alpha\beta}{{\cal H}_{\nu}}^{\alpha\beta}+ H_{\mu\alpha\beta}{H_{\nu}}^{\alpha\beta}\right )
-{{g_{\mu\nu}}\over{6}}\left ({\cal H}_{\alpha\beta\lambda}^2+H_{\alpha\beta\lambda}^2\right )\ .\label{EMS}
\end{eqnarray}
In particular a $4$-form in a higher dimensional gravity is known to generate a cosmological constant $\Lambda$$=$$-[(lv)^2/6]$$<$$0$ in GR \cite{dyn-const}. It governs a phantom field $\Psi$ in $d$$=$$5$ Einstein gravity and the trace of EMS tensor ensures a coupling $[-3l^2(\nabla\Psi)^2]$. 
On the other hand a bulk GT has been argued to source a boundary gravity \cite{Physica-NDS}. In $d$$=$$5$ bulk GT is sourced by the tr($T_{\mu\nu})$$=$$-[{\cal H}^2/3]$. A gauge theoretic torsion and a $4$-form respectively contribute topological corrections $(B_2$$ \wedge $$F_2)$ and 
$(\Phi$$ \wedge $${\cal F}_4)$ respectively in a boundary GR and redefines a NP gravity. 
A generic bulk/boundary correspondence ensures an equivalence between a bulk GT on $R^{1,1}$$ \otimes $$S_3$ and a boundary GR with a transverse dimension specified by the axionic scalar $\chi$. In principle a dynamical axion in bulk may add a topological coupling at the boundary GR on $R^{1,1}$$ \otimes $$S_2$.  Action may take a form: 
\be
S\rightarrow \int_{\partial\Sigma} \Big [{1\over{16\pi G_N}}\Big ( d^4x {{\sqrt{-g}}\ {\cal R}} -\ 4\pi B_2\wedge dA_1 \Big )\  + {1\over{4\pi\lambda^2}}{\cal H}_3\wedge d\chi\Big ]\ .\label{bGR}
\ee
The topological $(B_2$$ \wedge $$F_2)$ term in the action possesses its origin in a non-Newtonian potential within GR while the $({\cal H}_3$$ \wedge $$ F_1)$ term is sourced by a bulk GT. Their respective couplings signify their different origins. Generically both of them can count towards a NP correction to GR. However they donot modify the Einstein field equations of motion and hence all the exact geometries remain unaffected. Nonetheless they incorporate quantum corrections and they are believed to modify the topological characteristics of Riemannian geometries. It may be recalled under the bulk GT/boundary GR, that a Goldstone scalar in GT is assigned a vacuum expection value $<$$ \Phi $$>$ which in turn leads to a graviton in a boundary NP gravity. On the other hand the axionic scalar in bulk, under a change in space-time signature, turns out to be a ghost in the Einstein gravity. The ghost dynamics is cancelled by that of Goldstone scalar. It leaves behind an apparent $2$-form which leads to a theory of NP gravity. It reconfirms a decoupling of the axion (topological) in the boundary NP gravity. Then a decoupled bulk leads to a theory of NP gravity in $d$$=$$4$. It becomes by 
\bea
S=\int {1\over{16\pi G_N}}\Big ( d^4x {{\sqrt{-g}}\ {\cal R}} -\ 4\pi B_2\wedge dA_1 \Big )\ .\label{boundaryGR}
\eea
The topological action may be re-expressed as:
\bea
\int B_2\wedge F_2
&=&4\int\ d^4x{\sqrt{-g}}\ B_{\mu\nu}\nabla^{\mu}A^{\nu}\nonumber\\
&=&-4\int\ d^4x{\sqrt{-g}}\ A^2\ .\label{NP-gravity-1}
\eea
It assigns a mass term to a gauge field via topological coupling. Alternately a topological mass term may signify a dynamical gauge field hidden 
in an anti-braneworld within a gravitational pair. In particular a NP gravity is believed to describe a number of new phenomena \cite{EPJC-Priyabrat,NPB-Sunita, chen-IJTP} including (i) a multi RN black hole possibly underlying an tunneling instanton, (ii) a deep implication to the Big Bang cosmology and dark energy, (iii) degenerate Kerr vacua and (iv) landscape scenario in a type II superstring theories. However a detailed discussion on these topics is  beyond the scope of this paper.

\subsubsection{Gauge ansatz $\rightarrow$ Schwarzschild black hole}
Now we consider a $2$-form ansatz in bulk GT \cite{JHEP-Abhishek} and revisit the $d$$=$$4$ geometric perspective discussed in ref.\cite{PRD-Abhishek} with a renewed interest for the boundary gravity. A covariantly constant ${\cal B}_2$ and a dynamical $B_2$ ansatz for 
positive constants ($b,{\tilde P},P$) are given by 
\be
{\cal B}_{t\psi}=b={\cal B}_{R \psi}\ ,\quad B_{\theta\psi}= {\tilde P}\sin^2\psi \cot\theta\quad {\rm and}\quad B_{\psi\phi}=P\sin^{2}\psi \cos\theta\ .\label{ansatz-2}
\ee 
The anstaz consistently ensure the Coulomb gauge conditions (\ref{Coulomb}). The non-trivial components of GT becomes:
\be
{\cal H}_{\theta\phi\psi}={{bP}\over{l}}\sin^{2}\psi\sin\theta\qquad
{\rm and} \quad {\cal H}_{\theta\phi t}={\cal H}_{\theta\phi R}=\frac{-bPl}{R^{2}}\sin^{2}\psi \sin\theta\ .\label{ansatz-3}
\ee
They ensure that ${\tilde P}$ is a topological charge. Interestingly the dynamical $2$-form ansatz (\ref{ansatz-2}) is consistently governed by the Coulomb gauge (\ref{Coulomb}). In fact three local degrees of $2$-form in a perturbation theory, described with $\nabla_{\mu}$ derivative, are represented by one electric and two magnetic components (\ref{ansatz-3}) of GT. They are respectively given by
\be
{\mathbf{\tilde E}}=\left ({{bpl^2}\over{R^4}}, 0,0,0\right )\quad {\rm and}\qquad {\mathbf{\tilde M}}= \left ({{bpl}\over{R^3}}, {{bpl^2}\over{R^4}}, 0, 0\right )\ .\label{GT-EM}
\ee 
It shows that a higher form analogue of $EM$ field is sourced by a non-linear charge of GT. For a microscopic (small $R$) description, they turned out to be self-dual $i.e. |{\mathbf{\tilde E}}|$$=$$|{\mathbf{\tilde M}}|$. An electric non-linear charge due to a GT is a new phenomenon. Intuitively it may equivalently be viewed as a renormalized point charge in a cloud of photon. It helps to invoke the Heisenberg's uncertainty principle in an emergent gravity scenario.

\sp
\noindent
Generically an arbitrary EMS tensor, satisfying the continuity equation, is believed to source an exact geometric solution in Einstein gravity. Several attempts have been explored using various non-Abelian gauge theories in past. However a satisfactory gauge theoretic result leading to an educated guess for an EMS tensor is far from reality! In the recent past an attempt in this direction has partially been achieved with a $2$-form gauge theory \cite{JHEP-Abhishek}. The EMS tensor (\ref{EMS}) in $d$$=$$5$ bulk has been proposed to source a gravitational potential. An emergent metric has been shown to restore the gauge invariance in a GT. The metric in the braneworld scenario may be given by
\be
G_{\mu\nu}= \left (g_{\mu\nu}- {\cal B}_{\mu\alpha}{{\cal B}^{\alpha}{}}_{\nu} - l^2{\cal H}_{\mu\alpha\beta}{{\cal H}^{\alpha\beta}{}}_{\nu}\right )\ .\label{emergentM}
\ee 
The components have been worked out formally with an assigned spherical symmetry $S_3$ in this case. Interestingly a black hole line-element was 
approximated in a window via geometric engineering \cite{JHEP-Abhishek}. It becomes: 
\bea
&&ds^2=-\left (1-{{(lb)^2}\over{R^2}} + {{l^8b^2P^6}\over{R^8}} \right ) dt^2\ +\ 
\left ( 1 -{{(lb)^2}\over{R^2}} +{{l^8b^2P^6}\over{R^8}}\right )^{-1} dR^2\nonumber\\ 
&&\;\ + {{(lb)^2}\over{R^2}}\left (1 - {{(lP)^6}\over{R^6}} \right ) dt dR\ + {{2l^7bP^6}\over{R^6}}\left ( dt + dR \right )d\psi\
+ \left (1 - {{(lP)^6}\over{R^6}}\right )R^2 d\Omega^2_3\label{ads-3}
\eea
A vanishing torsion re-confirms a background black hole with a horizon radius $r_{h}$$=$$(lb)$. 
For small $R$, the braneworld scenario governs a microscopic black hole in $d$$=$$5$ defined with a GT coupling $\lambda=l^{3/2}$. 
It is re-assured by an empirical formula for a gravitational potential sourced by the ${\cal B}_{\mu\nu}$ field which in turn ensures a background geometry. 

\sp
\noindent
On the other hand a bulk GT, sourced by a conserved quantity $(bP)$, ensures non-linear electric and magnetic type fluctuations into the geometry. The emergent black hole turns out to be sourced by a self-dual $EM$ field underlying a GT. Presumably it provides a clue towards an eleven dimensional (torsion) theory. For large $R$, a non-linear magnetic fluctuation dominantly describes a macroscopic black hole. Generically a GT breaks the spherical symmetry in the emergent black hole(s). It describes a rotating (charged) black hole which is characterized by two horizons at $R_{\pm}$$=$$l(b\pm\delta P)$. It has been argued that a GT renormalizes a conserved charge of a background black hole to formally define a mass term at its horizon:
\be
2m=(lb)^2\left (1-{{(lP)^6}\over{r^6}}\right )_{r\rightarrow b}\ .\label{mass}
\ee
The off-diagonal terms, in the line-element (\ref{ads-3}), lead to intrinsic conserved quantities. They take opposite values on an anti-brane to that on a brane within a vacuum created gravitational pair. Thus the line-element (\ref{ads-3}), on a gravitational $(3{\bar 3})$-pair, has been argued to describe a  Schwarzschild black hole in a low energy limit \cite{JHEP-Abhishek}. The non-linearity (in a charge sourced by the GT) decouples in the limit and leads to an electric point charge. However the GT becomes dominantly magnetic and hence retains the non-linearity. In the limit the braneworld identifies with an exact vacuum solution in Einstein gravity underlying a formal correspondence between its couplings, $i.e.\ \lambda^2\rightarrow G_N$. Then a macroscopic black hole in a bulk GT is given by
\be
ds^{2}=-\left(1-{{2m}\over{R^2}}\right) dt^2+ \left(1-{{2m}\over{R^2}}\right)^{-1}dR^{2}+ R^2 d\Omega^2_3\ .
\ee
Recall that a propagating torsion (3-form potential) in GT requires a minimal $d$$=$$5$ 
which is similar to that of metric field dynamics which requires a minimal $d$$=$$4$ and defines GR.  Recall that an emergent metric $G_{\mu\nu}$ is a consequence of an explicit perturbative (defined with $\nabla_{\mu}$) gauge invariance under an $U(1)$ transformation of ${\cal B}_{\mu\nu}$ in GT theory \cite{JHEP-Abhishek,EPJC-Priyabrat}. Nonetheless the NP-term in the action uses the modified derivative ${\cal D}_{\mu}$ and an $U(1)$ gauge invariance is maintained for the GT. In other words the spontaneously broken (perturbative) $U(1)$ gauge invariance of ${\cal H}_3$ ensures massive ${\cal B}_{\mu\nu}$ which in turn defines an emergent metric at the expense of a mass of $2$-form. This observation is consistent with a fact that the local degrees of a massive $2$-form is precisely same as that of metric field in any space-time dimension. In the context it has been argued that the graviton in $d$$=$$4$ may equivalently be described by a massive $2$-form quantum \cite{PTEP-NS}. 

\section{Perihelion precession in bulk GT}
An emergent Schwarzschild metric in $d$$=$$5$ is characterized by one time-like Killing vector $K^{\mu}$$ \rightarrow $$(1,0,0,0,0)$ and six Killing vectors underlying the $S_3$. The translation symmetry in $\phi$ is characterized by $\xi^{\mu}$$ \rightarrow $$(0,0,0,0,1)$. Then the covariant Killing vectors are:
\be
K_{\mu}\rightarrow\left (-\left [1-\frac{2G_{N}M}{R^{2}}\right ], 0,\ 0,\ 0,\ 0\right )\quad {\rm and}\quad 
\xi_{\mu}\rightarrow \left (0,\ 0,\ 0,\ 0,\ R^2 \sin^{2}\psi \sin^{2}\theta\right )\ .\label{killing}
\ee
On an equatorial plane ($\psi$$=$${{\pi}\over2}$ and $\theta$$=$${{\pi}\over2})$, the conserved charges (energy $E$ and the magnitude of angular momentum $Q$) are given by
\be
E\rightarrow -K_{\mu}\frac{dx^{\mu}}{d\lambda}=\left (1-\frac{2G_{N}M}{R^2}\right ) \frac{dt}{d\lambda}\quad {\rm and}\quad Q\rightarrow \xi_{\mu}\frac{dx^{\mu}}{d\lambda}= R^2\frac{d\phi}{d\lambda}\ .\label{charge}
\ee
\begin{figure}[h]
\centering
\mbox{\includegraphics[width=0.7\linewidth, height=0.2\textheight]{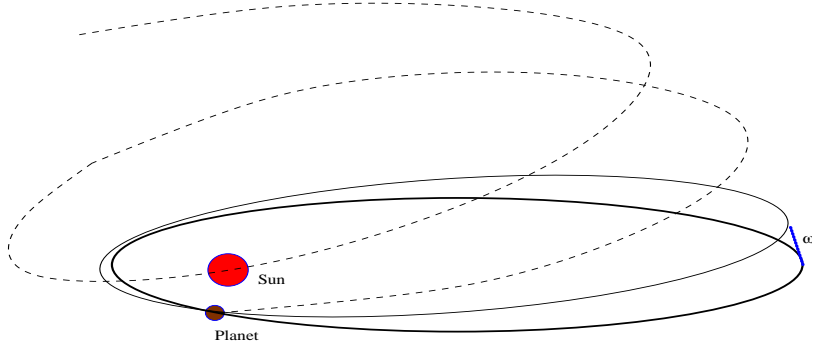}}
\caption{\it Schematic diagram shows a perihelion precession in a bulk GT for a planet around the Sun}
\end{figure}

\sp
\noindent
Newtonian gravity describes a circular orbit for a planet around the Sun. It becomes elliptical due to a planar effect of other massive bodies in the same plane. However the perihelion advances demonstrated in secton 4 re-confirms: $ \triangle$$ \phi $$=$$(6 \pi G_N)[M/Q]^2 $$ >0$. Presumably it 
reflects a deviation from a perfect elliptical orbit to an open path and hence signifies the role of GT in $d$$=$$5$ and for $d$$ > $$5$. Generically a constant of motion (\ref{motion-1}) in $d$$=$$5$ with Schwarzschild geodesics may take a form:
\bea
\left ({{dR}\over{d\lambda}}\right )^2&+&\left (1-{{2G_{N}M}\over{R^2}}\right )\left [1+R^2\left (\left [{{d\psi}\over{d\lambda}}\right ]^2 +\sin^2\psi \left [{{d\theta}\over{d\lambda}}\right ]^2\right )\right ]\nonumber\\
&&\qquad\qquad\qquad +\left ({{G_NQ^2}\over{R^2\sin^2\psi\sin^2\theta}}-\frac{2G_{N}^{2}MQ^2}{R^4\sin^2\psi\sin^2\theta}\right )=E^2\ .\label{GEnergy}
\eea
On an equatorial plane it may describe a classical particle of half-unit mass moving in one dimension \cite{arXiv-NRS}. It becomes 
\be
\left ({{dR}\over{d\lambda}}\right )^2 + \left (1-\frac{n}{R^2}-\frac{q^2}{R^{4}}\right )={\cal E}\ , \label{Energy}
\ee
where $n$$=$$(2G_{N}M$$-$$G_NQ^2)$, $q^2$$=$$({2G_{N}^{2}MQ^2})$ are constants and ${\cal E}$$=$$E^2$ is an analogue of total energy.
Interestingly the effective potential $V(R)$ in (\ref{Energy}) does not include a non-Newtonian term. It places a bulk GT in a different footing than GR. It may imply that a quantum correction to GR is likely to be governed in a bulk GT. In fact the $V_{\rm eff}$ in GT perturbation theory uses an emergent metric and hence both GT and GR are defined with a derivative $\nabla_{\mu}$ respectively in $d$$=$$5$ and $d$$=$$4$. This unusual energy analogue equation signifies that the actual motion involves the motion of a planet around the sun and hence $t(\lambda)$ and $\phi(\lambda)$ do join the $R(\lambda)$ equation (\ref{Energy}). Thus the actual scenario is drastically different from that of $d$$=$$1$ motion of a particle. The expression for $Q^2$ is used to re-express the eq(\ref{Energy}) under a change of variable. It takes a form:
\be
D_5=\left (\frac{dR}{d\phi}\right )^{2}+(1-{\cal E})\frac{R^4}{G_{N}Q^2}+\left (1-\frac{2M}{Q^2}\right )R^{2}-2G_{N}M =0\ .\label{Energy-1}
\ee
If $w$ is a fourth space coordinate then $R^2$$=$$(r^2$$+$$w^2)$. For $w^{2}$$ \ll $$r^{2}$, 
\be
R^4 \approx r^4\left ( 1+{{2w^2}\over{r^2}}\right )\qquad {\rm and}\quad \left ({{dR}\over{d\phi}}\right )^2\approx \left [1-{{w^2}\over{r^2}}\right ]\left ({{dr}\over{d\phi}}\right )^2\ .
\ee 
Then the  eq(\ref{Energy-1}) may be reduced to that in $d$$=$$4$. Using $X$$=$$Q^2(Mr)^{-1}$ the eq(\ref{Energy-1}) becomes 
\bea
{{d^2X}\over{d\phi^2}}&-& \left ({{3G_NM^2}\over{Q^2}}\right )X^2 + X=1,  \nonumber\\
{\rm and}\quad X^2\left (\frac{d^2X}{d\phi^2}\right )+X\left (\frac{dX}{d\phi}\right )^2&+&\tilde{N}X^3-\frac{3G_{N}Q^2}{w^2}X^2- NX={{Q^{4}}\over{M^{2}w}^{2}} \label{Energy-3} 
\eea
\bea
{\rm where}\quad\quad  N =2\left (1-{\cal E}\right )\frac{Q^2}{G_{N}M^2}-\frac{2Q^2}{Mw^{2}}\quad {\rm and}\quad \tilde{N}= \Big[\frac{4G_{N}M}{w^2}+\frac{4M}{Q^{2}}-2\Big]
\eea
The first differential equation identifies a solution (\ref{Eq-1D-5}) in GR. The second equation is differentiated w.r.t. $\phi$ and a further simplification leads to an equation:
\be
\left ({{15G_{N}M^2}\over{2Q^2}}\right )X^4\ + \left ( {{4G_{N}M}\over{w^2}}+\frac{4M}{Q^{2}}-4\right )X^3\ + {3\over2}\left ( 1-{{G_{N}Q^2}\over{w^2}} \right )X^2 \ + \left ({{Q^4}\over{2M^{2}w^{2}}} \right )=0\ .
\ee
The quartic equation may be re-expressed with two real roots ($\lambda$ and $\rho$). It is given by
\be
\left (X-\lambda\right )\left (X-\rho\right )\left (X+\left (\lambda+\rho\right )\right )\left (X+{{2Q^2}\over{15G_{N}M^2}}\Big({{4G_{N}M}\over{w^2}}+\frac{4M}{Q^{2}}-4\Big)\right )=0
\ee
A generic solution $X^{(GT)}$$=$$(X^{(GR)} + X_5)$ can be approximated with $w^2$$ \ll $$Q^2$ and identifying $\alpha=(3G_NM^2)Q^{-2}$ as the azimuthal precession in GR. Then: 
\be
w^2\approx G_{N}M\left (1+\frac{15}{8 \sqrt{3}}\frac{\sqrt{G_{N}}M}{Q}\right )\quad {\rm and}\;\ X_5\approx -\frac{4Q^2\alpha G_{N}}{3w^2}\ .\label{extra}
\ee
\vspace{-.15in}
\bea
{\rm Explicitly}\qquad \qquad\quad X^{(GT)}&=&1+e \cos\phi +e\alpha \phi \sin\phi+\alpha \tilde{\alpha}\ ,\quad \tilde{\alpha}=\left (1-\frac{4G_{N}Q^2}{3w^2}\right ),\nonumber\\
&=&1+e'\cos\phi+ e'' \alpha \phi\sin\phi\ ,\label{soln-1}
\eea
where 
$e'$$=$$(e+\alpha\tilde{\alpha}\cos\phi)$, and $e''\phi$$=$$(e\phi+ \tilde{\alpha}\sin\phi)$. For $w^2\rightarrow w_0^2={{4G_{N}Q^2}\over{3}}$, the $\tilde{\alpha}=0$ and $X_5$ contribution vanishes. Nevertheless for $\tilde{\alpha}\neq 0$:  
\bea
X^{(GT)}&=&1 +e' \cos\phi +e' \tilde{\alpha} \phi \sin \phi\ ,\ \nonumber\\
&=&1+e'\cos [(1-\tilde{\alpha})\phi]\ .\label{soln-2}
\eea
Interestingly the GT solution in a limit ${\tilde{\alpha}}$$ \rightarrow $$\alpha$ identifies with that in GR. This is due to a fact that 
$\alpha^2$ is insignificantly small. An estimate for $w$ is worked out for the motion of the planet Mercury around the Sun in torsion gravity. It yields $w=1.0\times 10^7 m$, where we have used $e=2\times 10^{-1}$, semi-major axis 
$a=5.8\times 10^{10}$m and $(G_NM_{\rm Sun}c^{-2})= 1.5\times 10^3$m. In the case an extra dimension turns out to be $10^3$ times smaller than the remaining three space dimensions which along with a time coordinate describes the GR \cite{arXiv-NRS}. Thus a perihelion advances by $w$ in an orthogonal direction to the remaining $3$-space coordinates. A small elevation in periodicity of the azimuthal angle $\phi$ is along a resultant direction to $w$ and $\mathbf{r}$. It is due to the non-planar effect underlying an intrinsic (non-commutative) nature of rotations which are only possible off a plane. Generically the bulk GT theory describes a spiral path for a planet and hence an open path! It is due to a propagating (axionic) scalar $\chi$ along the $w$-direction. Thus an assigned vacuum expectation $\chi_0$ can fine tune $w_0$ to a smaller value! A small $w_0$ ensures a nearly closed elliptical orbit in GR. A spiral path followed by a planet in GT may be approximated to describe an elliptical orbit on a slanted plane in GR.

\section{Perspectives of EM field in $d$$=$$5$ GT}
We recall the potential $V_q$ for a plausible physical interpretation in GR. 
It is obvious to note that the BF-term in the proposed action (\ref{q-action}) does not modify the known exact geometries in GR. Interestingly the $V_q$ may formally be identified with an electro-gravito (EG) or a magneto-gravito (MG) dipole term for $M$$\neq $$0$. The coined names presumably ensure the formation of an electric or magnetic dipole in presence of Einstein gravity. It may also be viewed through the coupling of vector field $A_{\mu}$ to the metric field $g_{\mu\nu}$ as in Einstein-Maxwell action. Thus two opposite EM charges $\pm Q$ are separated by $M$ and the EG or MG dipole is defined with a coupling $G_N$ which replaces the coulomb constant $(4\pi\epsilon_0)^{-1}$ in a typical electric dipole. However an electric or a magnetic dipole correction is ruled out in GR primarily due to a fact that EM charges are not sourced by the metric field. The Killing symmetries ensures that a dipole contribution varies as an inverse-cubic power of distance $r^{-3}$ from the dipole. It is insignificantly small for a large (length) scale when compared with the Newtonian (scalar) potential in GR. In particular all known exact solutions in GR and higher dimensional Einstein gravity do not include a $r^{-3}$ term in its geometry though the Killing symmetries ensure a dipole term in an effective potential. Nevertheless a quadruple may be configured with four EG or MG dipoles, with equal EM charge pair for each, placed to form a square such that the net charge at each vertex vanishes. Thus the vertices disappear to form a (gravitational or mass) loop, underlying a quadruple, and is known incorporate a correction to GR. 

\subsection{Role of a bulk $4$-form in GR}
Very recently the dipole potential has been shown to possess its origin in a bulk GT underlying a boundary GR by the authors \cite{arXiv-NRS}. It was argued that a dipole potential incorporates a topological correction to the GR and hence modifies the perihelion precession non-perturbatively. Furthermore a close inspection at the dipole term in the effective potential ensures its non-Newtonian origin and hence an exact solution in GR should exclude the dipole term. This is due to a fact that the coupling in GR is identified with the Newton's constant $G_N$. A dipole or a non-Newtonian term implies a non-scalar potential presumably leading to a $n$-form theory with an $U(1)$ gauge symmetry. Preliminary analysis reveals a mass dipole which may play a significant role in quantum cosmology. It may serve as a potential candidate to explore the origin of dark energy in universe. 

\sp
\noindent
An EG or a MG dipole in Einstein-Maxwell theory has been shown to govern by the $BF$-term. A topological number ensures windings which in turn would like to describe a multi Reissner-Nordstr$\ddot{\rm o}$m (RN) black hole. Thus a semi-classical vacua can be described with a quantum tunneling of an instanton via the $BF$-term underlying an EG or a MG dipole. The perspective of the boundary term has been shown to be sourced by the bulk $B_2$ dynamics \cite{Physica-NDS}. In fact the $B_2$ ansatz in the bulk gauge theory under $S_3$$\rightarrow $$S_2$, $i.e.$ for the second polar angle $\psi$$\rightarrow $${{\pi}\over2}$, has been worked out in ref\cite{PRD-Abhishek}. Our result matches with the expression for $V_q$ which sources an experimentally observed perihelion precession of planets in the solar system. Analysis may compel to revisit the perihelion precession with a renewed perspective in a bulk GT on $R^{1,1}$$ \otimes $$S_3$. 

\sp
\noindent
In the context we recall a $2$-form ansatz (\ref{ansatz-2}) to construct a geometric torsion (\ref{ansatz-3}) in a  perturbation theory. Thus, the gauge invariance is spontaneously broken in perturbation theory. As a result the 
${\cal H}_3$ may be treated as a gauge potential to define an $U(1)$ gauge invariant ${\cal F}_4$$=$$d{\cal H}_3$$ \neq $$0$. This in turn describes a propagating pseudo scalar presumably sourcing a gravitational instanton in the bulk GT. The components of field strength for a dynamical GT are worked out using the gauge ansatz (\ref{ansatz-3}). They are given by
\bea
{\cal F}_{t\psi \theta \phi}=-{\cal F}_{R \psi \theta \phi} = \frac{-2bP}{R^{2}}\sin 2\psi \sin \theta \quad {\rm and} \quad {\cal F}_{t R \theta \phi}=\frac{2bPl}{R^{3}}\sin^{2}\psi \sin \theta\ .\label{4-form}
\eea
On an equatorial (E) plane, the non-vanishing component of a $4$-form becomes
\be
{\cal F}_{t R\theta \phi}\rightarrow {{2bPl}\over{R^3}}\ .\label{equator}
\ee 
Under $R$$\rightarrow $$r$ an electric $4$-form identifies with a mass dipole term on an equatorial plance in GR. It is due to a fact that the spherical symmetry becomes insignificant on a causal plane. However the $3$-form equations of motion $\nabla^{\mu}{\cal F}_{\mu\nu\lambda\rho}$$=$$0$ in the perturbation GT theory ensure that the $4$-form ansatz does not contribute to the Newtonian force. It leads to a consistent geometric description in $d$$=$$5$. Thus the $4$-form ansatz re-confirms a holographic correspondence between a classical bulk (perturbation GT) and a boundary GR (with a NP-correction). It is consistent with the idea of ${\rm AdS}_5/{\rm CFT}_4$ correspondence \cite{susskind-hologram, witten-AdS,maldacena-AdS} which maps a weakly coupled bulk to a strongly coupled boundary. 

\sp
\noindent
Interestingly for $\psi$$=$$({\pi}/4)$ and $\theta$$=$$({\pi}/2)$, $i.e.$ a non-equatorial (N) plane, the $4$-form components become
\begin{eqnarray}
{\cal F}_{t\psi \theta \phi}=-{\cal F}_{R \psi \theta \phi}\rightarrow\frac{-2bP}{R^{2}}\quad {\rm and}\quad {\cal F}_{t R \theta \phi}\rightarrow\frac{2bPl}{R^{3}}\ .\label{4-form}
\end{eqnarray}
Thus a $4$-form contribution varies as inverse square distance ($R^{-2}$) on a non-equatorial plane. It is in addition to the inverse cubic distance 
($R^{-3}$) variation there. The bulk GT/boundary GR ensures that an $S_3\rightarrow S_2$ and hence a $4$-form may seen to contribute a potential term consistently described with the Newtonian gravity which is in addition to a dipole in GR. However the effective potential (\ref{GEnergy}) on the $E$ plane does not differ significantly from that on $N$ plane. This is evident with a trivial scaling $Q^2\rightarrow 2Q^2$. Importantly the effective potential on either (E or N) plane  does not incorporate a non-Newtonian term. It further re-confirms a classical description in $d$$=$$5$.

\sp
\noindent
A propagating GT and the metric in GR respectively require a minimal $d$$=$$5$ and $d$$=$$4$. In fact an emergent metric $G_{\mu\nu}$ has been shown to restore gauge invariance in GT theory \cite{JHEP-Abhishek}. Nonetheless the NP term in the action uses a modified derivative ${\cal D}_{\mu}$ and hence the $U(1)$ gauge invariance is maintained in GT. It has been argued that the graviton in $d$$=$$4$ may equivalently be described by the quantum of a massive $2$-form \cite{PTEP-NS}. 

\sp
\noindent
Now we revisit the perspective of an electromagnetic (EM) wave in $d$$=$$5$ bulk GT underlying a $2$-form theory with an $U(1)$ gauge symmetry. Needless to mention that a $2$-form gauge theory is Poincar$\acute{\rm e}$ dual to the $1$-form. In fact the duality symmetry holds good only with a gauge symmetry, $i.e.$ for mass-less forms. It implies that the propagating degrees of a $2$-form is equal to that of the $A_{\mu}$ field in the bulk. It is straightforward to observe that a mass term incorporates unequal number of additional local degrees to different forms and hence break the duality symmetry. Nonetheless a topological mass term does not incorporate any additional local degrees in the form theories and hence the duality symmetry is restored.

\subsection{$E$ field and transverse wave}
We begin with an electric point charge $q$ source in the bulk gauge theory. The gauge field ansatz in $d$$=$$5$ leads to a non-zero radial component $E_R$ of an electric field. They are respectively given by 
\be
A_{\mu}= -\frac{lq}{2R^2}\delta_{\mu t}\qquad {\rm and}\qquad F_{tR}={\mathbf{E}}=\left (-\frac{lq}{R^{3}},0,0,0\right )\ .\label{EM-ansatz} 
\ee
Then a non-vanishing component of the magnetic field ${\mathbf{M}}$ becomes
\bea
&&H_{\psi\theta\phi}={l\over2}{\sqrt{-g}}\ \varepsilon_{\psi\theta\phi tR}F^{tR}
=(lq)\sin^2\psi \sin\theta\ ,\quad {\rm where}\;\ \varepsilon_{tR\psi\theta\phi}=1\ ,\nonumber\\
{\rm and}\;&&{\mathbf{M}}=\left ({{lq}\over{R^3}},0,0,0\right )\ .\label{M-field}
\eea
It ensures the self duality $|{\mathbf{E}}|$$=$$|{\mathbf{M}}|$ in $d$$=$$5$. Furthermore an electric field (\ref{EM-ansatz}) alone is known to generate a magnetic field in presence of a GT \cite{JHEP-Abhishek} as the EM field has been shown to receive a correction. In particular the $2$-form(s) ansatz (\ref{ansatz-2}) leading to a GT (\ref{ansatz-3}) explores a theoretical feasibility only for a magnetic monopole $q_m$ from  an electric charge $q$.  Remarkably a GT does not seem to allow a reverse process involving a generation of $q$ from $q_m$. 
The EM field in GT is given by
\bea
&&{\cal F}_{\mu\nu}=F_{\mu\nu} + {{\cal H}_{\mu\nu}{}}^{\alpha}A_{\alpha}\nonumber\\
{\rm or}&&{\cal F}_{tR}=F_{tR}\quad {\rm and}\quad {\cal F}_{\theta\phi}={{\cal H}_{\theta\phi}{}}^tA_t=- {{lq_m}\over{R^4}} \sin^2\psi\sin\theta\ , \label{M-charge}
\eea
where $q_m$$=$$[(lqbp)/2]$. The phenomenon is in agreement with an experimental fact that magnetic monopole has not been found. It is indeed a theoretical artifact of a GT and may place the bulk GT leading to a boundary gravity on a prominent edge. Though the $E$-field is linear, the $M$-field turns out to be non-linear due to its inherent coupling to the GT. The later becomes gauge invariant with ${\cal F}_{\theta\phi}^2$ and is given by
\be
g^{\theta\theta}g^{\phi\phi}{\cal F}_{\theta\phi}{\cal F}_{\theta\phi}={{q_m^2}\over{R^{12}}}\ .\label{Magnetic-nl}
\ee
Then the $E$ field and $M$ field in the case turns out to be described by 
\be
{\mathbf{E}}\rightarrow\left (-{{lq}\over{R^3}},0,0,0\right )\qquad {\rm and}\qquad {\mathbf{M}}\rightarrow \left ({{l^3q_m}\over{R^6}},0,0,0\right )\ .\label{EM-bulk} 
\ee
They show that the generated magnetic field significantly dominants over the electric field for small $R$. It ensures a fact that generically a GT is a high energy phenomenon defined with an UV cut-off. In bulk GT, the magnitude $|{\mathbf{M}}|$ depends on $|{\mathbf{E}}|$. However both ${\mathbf{M}}$ and ${\mathbf{E}}$ are treated independent as they are oriented along different  directions. In fact they characterize two polarizations of EM theory in $d$$=$$5$ bulk and have been argued to govern an apparent spin $2$ (mass-less) $2$-form in subsection 3.2 which turns out to be a Poincar$\acute{\rm e}$ dual description to the $A_{\mu}$. The apparent $2$-form may also be viewed in terms of a massive $2$-form whose one local degree is cancelled by that of a GT. This in turn ensures a decoupling of the GT and hence an apparent $2$-form turns out to be linear. Thus a GT description (\ref{torsion-bulk}) in a decoupling limit may seen to describe an apparent $2$-form 
in addition to a (spin zero) Goldstone scalar in an $U(1)$ gauge theory. Arguably an absorption of the Goldstone scalar by the apparent $2$-form leads to $6$ local degrees and hence a massive $2$-form in $d$$=$$5$. 

\sp
\noindent
A wave vector ${\mathbf{k}}$ in the $d$$=$$5$ bulk EM description, underlying an apparent $2$-form, ensures its transverse nature and is  described with a periodicity of $2\pi$. Intuitively a superposition of an oscillating circular wave (sourced by Goldstone scalar) along with the bulk EM wave may lead to a transverse propagation of a group of oscillating circular waves with a reduced periodicity of $\pi$. See schematic diagrams in figure-2,3 and 4. An empirical formula $({\rm spin}$$=$$2\pi [{\rm periodicity}]^{-1})$ further assigns a spin $2$ presumably in a quantum description to a profile of transverse waves in bulk. Alternately the scenario may be viewed in terms of four EG (or MG) dipoles forming an aprior square with opposite polarity at each vertex. A loop so produced is described purely by the mass points. Hence a massive test particle in motion in the same plane to that of the loop would likely to change the shape of the loop back and forth from a circle to an ellipse due to the Newtonian gravity. Their transverse propagation is believed to govern the gravitational wave in a boundary GR. Interestingly the transverse nature of gravitational wave is not an assumption (as considered in a linearized GR) but is intrinsic to a bulk GT. It may further putforward the bulk GT/boundary GR correspondence to its merit. 

\sp
\noindent
On the hand a generic bulk GT description (\ref{torsion-bulk}) ensures an oscillating spiral wave profile along a nearly transverse direction naively underlying a massive $2$-form and a mass-less $3$-form (GT) dynamics. The spiral profile may ensure a spin $2$ in an appropriate quantum theory. A spiral profile of waves may also be imagined via a nearly transverse motion of two parallel EG (or MG) dipoles where the second dipole undergoes a Coulomb repulsion by the first along a nearly circular or an elliptical path. In a GT (local degree) decoupling limit the spiral (open) profile identifies with that of a  circular/elliptical (closed loop) and retains the transverse nature. 

\begin{figure}[h]
\centering
\mbox{\includegraphics[width=0.6\linewidth, height=0.17\textheight]{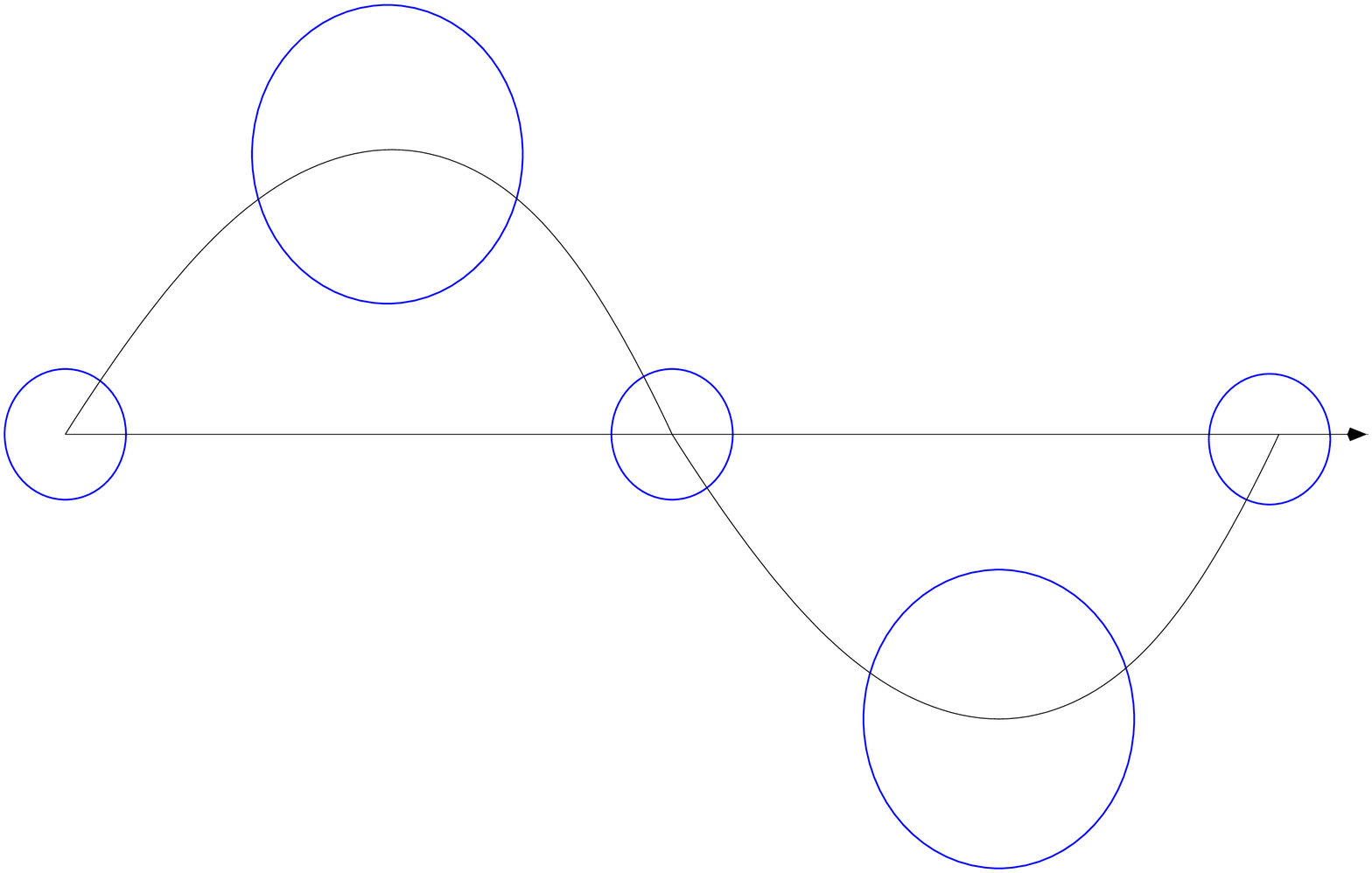}}
\caption{\it Schematic  superposition of a circular wave and a transverse EM wave}
\end{figure}

 \begin{figure}[h]
\centering
\mbox{\includegraphics[width=0.6\linewidth, height=0.16\textheight]{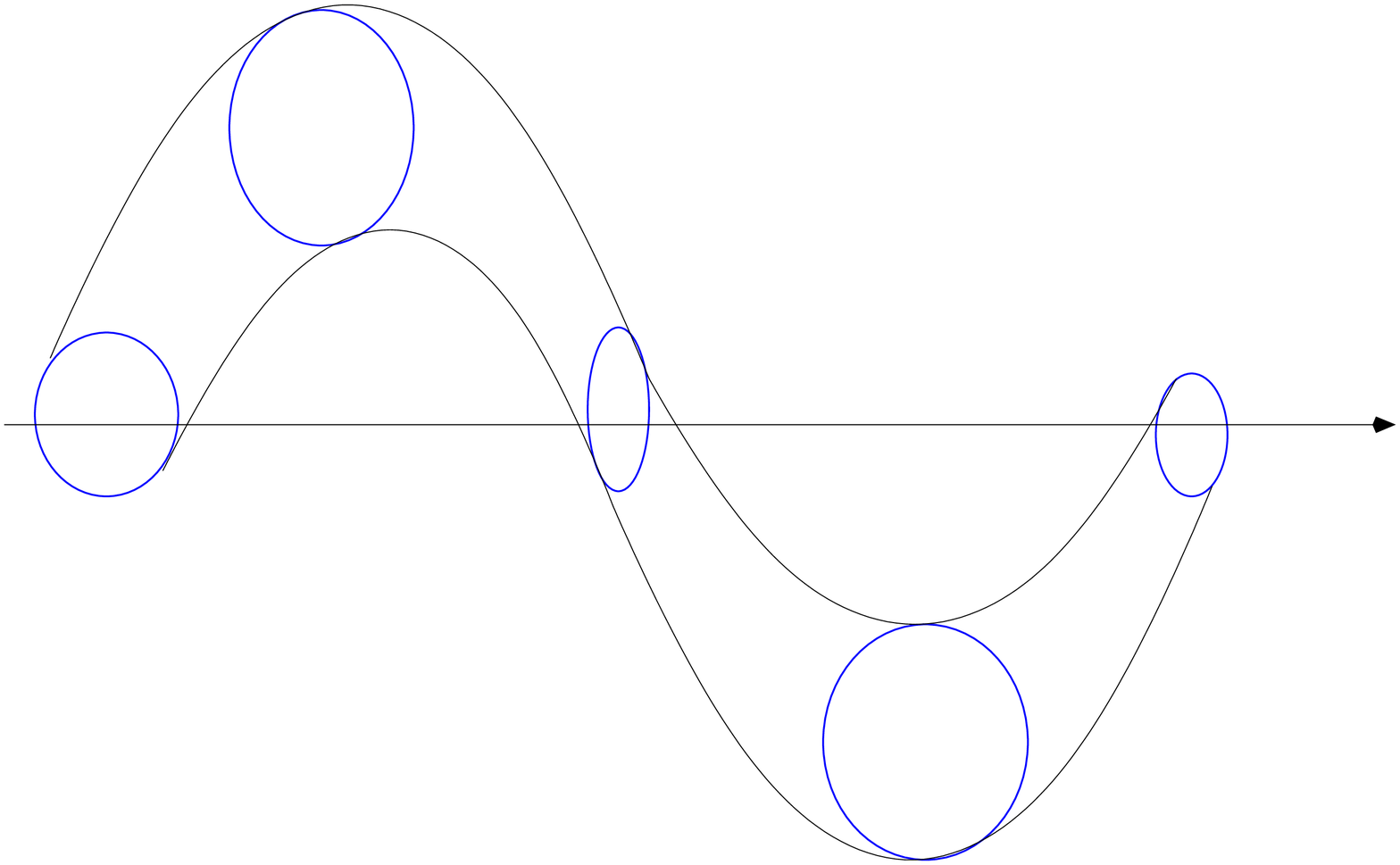}}
\caption{\it Schematic superimposed transverse wave profile in bulk EM theory}
\end{figure}

\begin{figure}[h]
\centering
\mbox{\includegraphics[width=0.6\linewidth, height=0.17\textheight]{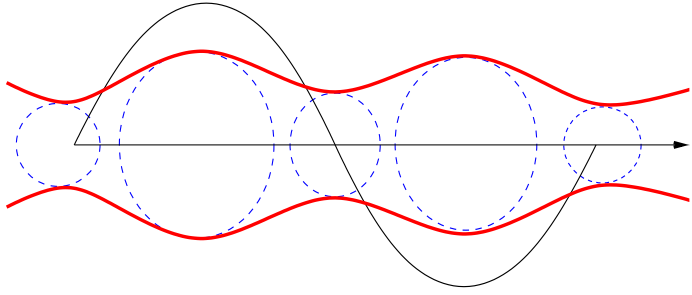}}
\caption{\it Schematically  a reduced half wavelength governs a spin $2$ particle in a dual scenario.}
\end{figure}

\subsection{Gravitational wave/particle duality}
The bulk GT $\leftrightarrow$ boundary GR proposal underlie a correspondence between a perturbative or weakly coupled gauge theory and a NP or strongly coupled gravity theory. In fact a NP correction has been shown to be sourced by a non-trivial GT dynamics ${\cal F}_4$$=$$d{\cal H}_3$. In a torsion decoupling limit, $i.e.$ for a non-propagating GT, the local degrees in the bulk has been identified with a boundary GR. Interestingly the decoupling limit has been argued to decouple the bulk non-linearity which in turn is believed to describe a gravitational wave in a boundary description. The wave profile is known to describe a propagating loop of varying shapes orthogonal to each other, $i.e.$ from circle $\rightarrow$ vertical ellipse $\rightarrow$ circle $\rightarrow$ horizontal ellipse and so on. Thus a gravitational wave is described by an envelope of these varying shapes propagating in a perpendicular direction to the loop. These group of waves are defined with a reduced periodicity $\pi$ which in turn empirically ensure a spin $2$ particle  in a quantum description.

\sp
\noindent
At this juncture we recall that an exact soloution in GR underlies the Riemannian geometry. Three limits (week gravity, stationary and non-relativistic) together in GR leads to the Newtonian gravity. It is the Newtonian potential which prohibits a quantum description! This may be viewed through a fact that Heisenberg's uncertainty principle is not compatible with the point masses which define the Newtonian gravity. It would make the conjugate momentum infinite as the conserved force turns out to be an infinite range. Though the issue appears similar to that of Coulomb force in EM theory it differs significantly in a quantum theory. An $U(1)$ gauge theoretic description shields the charge (say electron with a photon cloud) via re-normalization and hence the conjugate momentum remains finite. Alternately a notion of an extended or non-linear (conserved) charge automatically sets up a lower cut-off in the length scale and hence an UV cut-off in a relativistic quantum description. A non-linear charge is known to influence the global properties without any change in its local properties. Without compromising the characteristics, the GR can never lead to a consistent quantum (metric) field theoretic description. 

\sp
\noindent
In addition an interacting nature in-built in Newtonian gravity does not allow a free field theory description and hence rules out a perturbative perspective in GR. Thus a free graviton is an idealistic realization and may serve as an academic exercise but it would not completely justify a metric quantum. A consistent quantum theory of gravity would require a non-perturbative (NP) formulation. It may imply that a NP correction should not include the Newtonian potential and hence is not within the GR but from outside. This in turn enforces a background independent correction. It would like to bring-in a drastically different perception to quantum gravity than that in a quantum field theory (QFT) or in a gauge theory. Along the line various shades of quantum gravity may be revisited. For instance we recall a quantum effect in a special case of Einstein gravity in $(2$$+$$1)$-dimensions leading to the BTZ black hole \cite{BTZ}. It has been shown that a number of discrete values of a conserved charge in BTZ geometries ensure ${\rm AdS}_3$ bound states. Thus a continuum description within an ${\rm AdS}_3$ bound state is separated with a series of bare singularities from another bound state. These singularities may be avoided with a plausible tunneling effect between the ${\rm AdS}_3$ bound states. Similarly, but in a different context, a large degeneracy sourced by a non-linear charge has been shown to arise in a quantum (gauge theoretic) description leading to an emergent metric \cite{NPB-Sunita}. The degeneracy was shown to disappear to describe the Kerr-Newman black hole in a low energy limit.

\sp
\noindent
However a possible quantization of gravity happens to be enforced at least by a: (i) number of experimental data in observational cosmology, (ii) theoretical perspective in high energy physics and (iii) conceptual realization of some observed phenomena. In the context a higher ($d$$>$$4$) dimensional (Einstein gravity) classical description is believed to incorporate some quantum effects into the GR through gauge field couplings and is known to describe a semi-classical theory. However a complete quantum theory of gravity is lacking. 

\sp
\noindent
Interestingly closed superstring theories in $d$$=$$10$ have been known to describe a graviton in addition to the mass-less field quanta (dilaton and $2$-form) and a large number of massive quanta in its spectrum \cite{GSW}. For instance the Polyakov action for a closed bosonic string essentially ensures a free string propagation described with a cylindrical worldsheet. A canonical quantization on the string world-sheet leads to a non-interacting graviton dynamics on $R\otimes S^1$ topology. A free graviton theory presumably ensure the perturbtion perspective prominent with a generalization of the Minkowski metric to an arbitrary background in a non-linear sigma model string world-sheet action. The essential theme is in agreement with a free QFT which in turn ensures the perturbative Feynman diagrams for an interaction process. Along the line  a free (metric) field theory is also ruled out. It  further re-ensures a NP formulation for gravity as a perturbative GR would break down.

\sp
\noindent
This in turn assigns a spiral or open path to a planet around the sun and hence gives rise to an advancement of the perihelion. Interestingly two ends of an open path may imagined to be connected with a dipole in a boundary GR. This in turn is approximated to yield a nearly closed elliptical loop and hence a metric description in GR. Alternately a decoupling limit, for a propagating GT, has been argued to source a NP gravity, $i.e.$ a metric field dynamics along with a topological correction. It provokes thought to believe that a gravitational wave profile underlying two transverse polarizations in GR in a NP gravity would be described by a cylinder topology. Needless to mention that a topological correction is background invariant and hence all closed loops are equivalent to each other. The wave nature disappears as the periodicity is removed by an intrinsic topological correction in GR. Interestingly the cylindrical topology may formally be identified with a graviton underlying a free closed string propagation. Remarkably analysis reveals a graviton in a dual scenario to the gravitational waves. An analogy with the wave/particle duality in quantum mechanics further validates our correspondence between a bulk GT and boundary NP gravity. 

\sp
\noindent
Intuitively a non-Newtonian potential in GR  may not incorporate an interaction due to masses!  As a result a massive test particle motion on the same plane to a loop in a non-Newtonian gravity would not change the shape and size of the propagating loop. It may be realized with a free closed string propagation and hence a graviton and a dilaton in its energy spectrum. Thus a weak gravity sources a gravitational wave in a classical theory and may describe a non-interacting graviton in a NP gravity.

\sp
\noindent
 On the other hand GR evolves with a space-time curvature and describes a macroscopic Schwarzschild black hole. A topological correction to GR would like to describe an interacting graviton in a NP gravity. A preliminary analysis with first three terms in effective potential (\ref{motion-2}) may ensure a Reissner-Nordstrom (RN) black hole. A mass dipole correction sourced by the fourth term in effective potential would like to modify the topological characteristics of RN geometry. On the other hand the dipole from a bulk perspective ensures an instanton correction to the RN geometry. Thus the NP gravity may presumably lead to a multi RN black hole with a tunneling instanton. A detailed analysis is beyond the scope of this paper and is in progress.

\section{Concluding remarks}
We began this paper with a list of  evidences to enlighten our proposed equivalence between a bulk GT/boundary GR \cite{Physica-NDS}. It was argued that a weakly coupled bulk dynamics primarily described by a $2$-form may equivalently be governed by a strongly coupled metric dynamics in boundary. Interestingly a topological coupling of the form $(B_2\wedge F_2)$ in the boundary GR was shown to govern a NP gravity. The NP coupling was shown to be sourced by a $4$-form field strength underlying a propagating GT in bulk. Most importantly a NP coupling was shown to be sourced by a non-Newtonian potential predicted by the isometries in GR. The coupling was realized in terms of a dipole correction to GR. In the context we have performed a rigorous analysis to compute the advances in a perihelion in bulk GT. It was shown that an advancement in azimuthal angle non-perturbatively modifies that in GR underlying an instanton effect.

\sp
\noindent
Interestingly the gravitational wave/particle duality in $d$$=$$4$ NP gravity was argued to formally identify a graviton by taking an analogy from a free closed superstring theory in $d$$=$$10$. The extra $6$ space dimensions in superstring theory, when compared with that of a NP gravity, may ensure a much lower energy then the Planck scale to the latter. However both the formulations share a minimal length scale though they turn out to be of different order of magnitude. We re-iterate that a minimal scale in NP gravity is incorporated by the topological windings sourced by a non-Newtonian potential. Thus a source underlying a quantum correction may be described by a non-scalar field such as non-zero form fields or generically a coupling of a form field with another. Interestingly a non-scalar field operation on the commutator of covariant derivatives (\ref{commutator}) in GR ensures a non-vanishing curvature. It adds to the Riemannian curvature. Together they lead to a higher energy (boundary) NP gravity underlying a propagating torsion in bulk. 

\sp
\noindent
Furthermore a renewed perspective of an EM wave in $d$$=$$5$ bulk was worked out in a $2$-form theory with an $U(1)$ gauge symmetry. Inspite of an odd dimension the Poincare duality with a $2$-form ansatz was shown to satisfy the self-duality between an electric and a magnetic field vector in bulk. Remarkably a bulk GT ensured an extended magnetic charge sourced by an electric point charge which is in agreement with the present day experimental observation for no magnetic monopole. However ${\mathbf{E}}$ and ${\mathbf{B}}$ vectors turn out to be independent in the bulk and they were argued to describe two polarizations of an EM wave sourced by an apparent $2$-form. Computation for local degrees ensured that the apparent $2$-form in bulk corresponds to a graviton in the boundary NP gravity. Thus a bulk GT/boundary GR correspondence automatically assigns a transverse nature to the gravitational wave which is otherwise an assumption in a linearized approximation to GR.

\def\anp{Ann. of Phys.}
\def\prl{Phys.Rev.Lett.}
\def\prd#1{{Phys.Rev.}{\bf D#1}}
\def\jhep{JHEP\ {}}{}\noindent
\def\cqg{{Class.\& Quant.Grav.}}
\def\plb#1{{Phys. Lett.} {\bf B#1}}
\def\npb#1{{Nucl. Phys.} {\bf B#1}}
\def\mpl#1{{Mod. Phys. Lett} {\bf A#1}}
\def\ijmpa#1{{Int.J.Mod.Phys.}{\bf A#1}}
\def\mpla#1{{Mod.Phys.Lett.}{\bf A#1}}
\def\rmp#1{{Rev. Mod. Phys.} {\bf 68#1}}
\def\jmp{J.Math.Phys.}
\def\cmp{Commun.Math.Phys. }
\def\epj#1{{Eur.Phys.J. }{\bf C #1}}
\def\ijmpd#1{{Int.J.Mod.Phys.}{\bf D#1}}
\def\physica{Phys. Scripta }
\def\ptp{ Prog.Theo.Phys.}
\def\ijtp{Int. J. Theo. Phys.}

\end{document}